%% file: kapand.tex
\documentclass[twocolumn]{aastex61}
\usepackage{amssymb,amsmath}

\usepackage{wasysym}

\shorttitle{$\kappa$ And \lowercase{b}}
\shortauthors{Currie et al.}
\begin{document}
\title{SCExAO/CHARIS Near-Infrared Direct Imaging, Spectroscopy, and Forward-Modeling of $\kappa$ And \lowercase{b}: A Likely Young, Low-Gravity Superjovian Companion}
\correspondingauthor{Thayne Currie}
\email{thayne.m.currie@nasa.gov,currie@naoj.org}
\author{Thayne Currie}
\affiliation{NASA-Ames Research Center, Moffett Field, CA, USA}
\affiliation{Subaru Telescope, National Astronomical Observatory of Japan, 
650 North A`oh$\bar{o}$k$\bar{u}$ Place, Hilo, HI  96720, USA}
\affiliation{Eureka Scientific, 2452 Delmer Street Suite 100, Oakland, CA, USA}
\author{Timothy D. Brandt}
\affiliation{Department of Physics, University of California, Santa Barbara, Santa Barbara, California, USA}
\author{Taichi Uyama}
\affiliation{Department of Astronomy, The University of Tokyo, 7-3-1, Hongo, Bunkyo-ku, Tokyo 113-0033, Japan}
\author{Eric L. Nielsen}
\affiliation{Kavli Institute for Particle Astrophysics and Cosmology, Stanford University, Stanford, CA 94305, USA}
\author{Sarah Blunt}
\affiliation{Department of Astronomy, Harvard University, 60 Garden St., Cambridge, Massachusetts, USA}
\author{Olivier Guyon}
\affiliation{Subaru Telescope, National Astronomical Observatory of Japan, 
650 North A`oh$\bar{o}$k$\bar{u}$ Place, Hilo, HI  96720, USA}
\affil{Steward Observatory, University of Arizona, Tucson, AZ 85721, USA}
\affil{College of Optical Sciences, University of Arizona, Tucson, AZ 85721, USA}
\affil{Astrobiology Center of NINS, 2-21-1, Osawa, Mitaka, Tokyo, 181-8588, Japan}
\author{Motohide Tamura}
\affiliation{Department of Astronomy, The University of Tokyo, 7-3-1, Hongo, Bunkyo-ku, Tokyo 113-0033, Japan}
\affiliation{Astrobiology Center of NINS, 2-21-1, Osawa, Mitaka, Tokyo, 181-8588, Japan}
\author{Christian Marois}
\affiliation{National Research Council of Canada Herzberg, 5071 West Saanich Rd, Victoria, BC, V9E 2E7, Canada}
\affiliation{University of Victoria, 3800 Finnerty Rd, Victoria, BC, V8P 5C2, Canada}
\author{Kyle Mede}
\affiliation{Department of Astronomy, The University of Tokyo, 7-3-1, Hongo, Bunkyo-ku, Tokyo 113-0033, Japan}
\author{Masayuki Kuzuhara}
\affiliation{Astrobiology Center of NINS, 2-21-1, Osawa, Mitaka, Tokyo, 181-8588, Japan}
\author{Tyler D. Groff}
\affiliation{NASA-Goddard Space Flight Center, Greenbelt, MD, USA}
\author{Nemanja Jovanovic}
\affiliation{Department of Astronomy, California Institute of Technology, 1200 E. California Blvd., Pasadena, CA 91125, USA}
\author{N. Jeremy Kasdin}
\affiliation{Department of Mechanical Engineering, Princeton University, Princeton, NJ, USA}
\author{Julien Lozi}
\affiliation{Subaru Telescope, National Astronomical Observatory of Japan, 
650 North A`oh$\bar{o}$k$\bar{u}$ Place, Hilo, HI  96720, USA}
\author{Klaus Hodapp}
\affiliation{Institute for Astronomy, University of Hawaii,
640 North A`oh$\bar{o}$k$\bar{u}$ Place, Hilo, HI 96720, USA}
\author{Jeffrey Chilcote}
\affiliation{Department of Physics, University of Notre Dame, 225 Nieuwland Science Hall, Notre Dame,
IN, 46556, USA}
\author{Joseph Carson}
\affiliation{Department of Physics, College of Charleston}
\author{Frantz Martinache}
\affiliation{Universit\'{e} C\^{o}te d'Azur, Observatoire de la C\^{o}te d'Azur, CNRS, Laboratoire Lagrange, France}
\author{Sean Goebel}
\affiliation{Institute for Astronomy, University of Hawaii,
640 North A`oh$\bar{o}$k$\bar{u}$ Place, Hilo, HI 96720, USA}
\author{Carol Grady}
\affiliation{Eureka Scientific, 2452 Delmer Street Suite 100, Oakland, CA, USA}
\affiliation{NASA-Goddard Space Flight Center, Greenbelt, MD, USA}
\author{Michael McElwain}
\affiliation{NASA-Goddard Space Flight Center, Greenbelt, MD, USA}
\author{Eiji Akiyama}
\affiliation{Hokkaido University, Department of Cosmosciences, Hokkaido University, Kita-ku, Sapporo, Hokkaido 060-0810, Japan}
\author{Ruben Asensio-Torres}
\affiliation{Department of Astronomy, Stockholm University, AlbaNova University Center, SE-106 91 Stockholm, Sweden}
\author{Masa Hayashi}
\affiliation{National Astronomical Observatory of Japan, 2-21-2, Osawa, Mitaka, Tokyo 181-8588, Japan}
\author{Markus Janson}
\affiliation{Department of Astronomy, Stockholm University, AlbaNova University Center, SE-106 91 Stockholm, Sweden}
\author{Gillian R. Knapp}
\affiliation{Department of Astrophysical Sciences, Princeton University, Princeton, New Jersey, USA}
\author{Jungmi Kwon}
\affiliation{ISAS/JAXA, 3-1-1 Yoshinodai, Chuo-ku, Sagamihara, Kanagawa 252-5210, Japan}
\author{Jun Nishikawa}
\affiliation{National Astronomical Observatory of Japan, 2-21-2 Osawa, Mitaka, Tokyo 181-8588, Japan}
\author{Daehyeon Oh}
\affiliation{National Meteorological Satellite Center, 64-18 Guam-gil, Gawnghyewon-myeon, Jincheon-gun, Chungcheongbuk-do 27803, Republic of Korea}
\author{Joshua Schlieder}
\affiliation{NASA-Goddard Space Flight Center, Greenbelt, MD, USA}
\author{Eugene Serabyn}
\affiliation{Jet Propulsion Laboratory, California Institute of Technology, Pasadena, CA, 171-113, USA}
\author{Michael Sitko}
\affiliation{Center for Extrasolar Planetary Systems, Space Science Institute, 1120 Paxton Ave., Cincinnati, OH 45208, USA}
\author{Nour Skaf}
\affiliation{Institut d?Optique Graduate School, Université Paris-Saclay, Paris, France}

\begin{abstract}
We present SCExAO/CHARIS high-contrast imaging/$JHK$ integral field spectroscopy of $\kappa$ And b, a directly-imaged low-mass companion orbiting a nearby B9V star.   We 
detect $\kappa$ And b at a high signal-to-noise and extract high precision spectrophotometry using a new forward-modeling algorithm for (A-)LOCI complementary to KLIP-FM developed by Pueyo et al. (2016).  
$\kappa$ And b's spectrum best resembles that of a low-gravity L0--L1 dwarf (L0--L1$\gamma$). 
Its spectrum and luminosity are very well matched by 2MASSJ0141-4633 and several other 12.5--15 $M_{\rm J}$ free floating members of the 40 $Myr$-old Tuc-Hor Association, 
 consistent with a system age derived from recent interferometric results for the primary, a companion mass at/near the deuterium-burning limit (13$^{+12}_{-2}$ M$_{\rm J}$), and a companion-to-primary mass ratio characteristic of other directly-imaged planets  ($q$ $\sim$ 0.005$^{+0.005}_{-0.001}$).    We did not unambiguously identify additional, more closely-orbiting companions brighter and more massive than $\kappa$ And b down to $\rho$ $\sim$ 0\farcs{}3 (15 au).    
 SCExAO/CHARIS and complementary Keck/NIRC2 astrometric points reveal clockwise orbital motion.   Modeling points towards a likely eccentric orbit: a subset of acceptable orbits include those that are aligned with the star's rotation axis.   However, $\kappa$ And b's semimajor axis is plausibly larger than 75 au and in a region where disk instability could form massive companions.

Deeper $\kappa$ And high-contrast imaging and low-resolution spectroscopy from extreme AO systems like SCExAO/CHARIS and higher resolution spectroscopy from Keck/OSIRIS or, later, IRIS on the \textit{Thirty Meter Telescope} could help clarify $\kappa$ And b's chemistry and whether its spectrum provides an insight into its formation environment.
\end{abstract}
\keywords{planetary systems, stars: early-type, stars: individual:  HD 222439 -- stars: individual (HD 222439), techniques: high angular resolution} 
\section{Introduction}
The past decade of facility high-contrast imaging systems and now dedicated \textit{extreme} adaptive optics-based planet imagers have revealed the first direct detections of planets around nearby, young stars \citep{Marois2008,Marois2010a,Lagrange2010, Kuzuhara2013, Carson2013, Quanz2013,Rameau2013,Currie2014a,Currie2015a,Macintosh2015,Chauvin2017,Keppler2018}.    Their range of masses (2--15 $M_{\rm J}$) and orbital separations (10--150 au) make them key probes of jovian planet formation models \citep[e.g.][]{Boss1997,KenyonBromley2009}.  The companions' photometry reveal clear differences with field brown dwarfs and evidence for extremely cloudy and/or dusty atmospheres
\citep{Currie2011a}.   

Integral field spectrographs (IFS) further clarify the atmospheric properties of young planet-mass companions, revealing 
tell-tale signs of low surface gravity from sharper, more point-like $H$-band peaks \citep[e.g.][]{Barman2011,AllersLiu2013}.  Hotter, early L type planets at very young ages (1--10 $Myr$) may also exhibit a red, rising slope through $K$-band, also a sign of low surface gravity \citep{Canty2013,Currie2014a}.    While the near-infrared (near-IR) spectra of some cooler L/T and T-type directly-imaged planets show evidence for more extreme clouds, more vigorous chemical mixing, and/or lower gravities than found in (nearly all of) even the youngest, lowest mass objects formed by cloud fragmentation \citep[e.g.][]{Currie2011a,Bonnefoy2016, Rajan2017, Chauvin2018}, L-type young directly-imaged planets can be nearly indistinguishable from free floating, planet-mass analogues with identical ages \citep[e.g.][]{AllersLiu2013,Chilcote2017, Dupuy2018}.

The directly-imaged low-mass companion to the B9V star $\kappa$ Andromedae \citep[$\kappa$ And b;][]{Carson2013} is an object whose properties could be clarified by new, high-quality IFS data.    Based on $\kappa$ And b's luminosity and the primary's proposed status as a sibling of HR 8799 in the 30--40 $Myr$-old Columba association \citep{Zuckerman2011,Bell2015}, \citeauthor{Carson2013} estimated its mass to be 12.8 $M_{\rm J}$.  
 Using broadband photometry, \citet{Bonnefoy2014a} suggest a spectral type of M9--L3 and find some evidence for photospheric dust but fail to constrain $\kappa$ And b's surface gravity and admit a wider range of possible ages and thus masses.  The Project 1640 IFS-based follow-up study by \citet{Hinkley2013} question whether $\kappa$ And is a Columba member, derive a much older age of 220 $Myr$, and argue that $\kappa$ And b's spectrum suggests the companion is not planetary mass.   
   However, subsequent studies based on the primary admit the possibility that the system is young ($t$ $\sim$ 30--40 $Myr$; and thus the companion could be low mass) \citep{DavidHillenbrand2015,BrandtHuang2015}.   Furthermore,  CHARA interferometry precisely constraining the rotation rate, gravity, temperature, and luminosity and comparing these properties to stellar evolution models favor a young age \citep{Jones2016}.   New, higher quality IFS data for $\kappa$ And b can better clarify whether the companion shares properties (e.g. surface gravity) more similar to the young planet-mass objects or older, deuterium burning brown dwarfs.

In this study, we report new $JHK$ direct imaging and spectroscopy of $\kappa$ And b obtained with the \textit{Subaru Coronagraphic Extreme Adaptive Optics} project coupled to the CHARIS integral field spectrograph \citep{Jovanovic2015a,Groff2013}.  We analyze these data and combine them with archival Keck/NIRC2 imaging to yield new constraints on $\kappa$ And b's atmosphere and orbit.   

\section{SCExAO/CHARIS Data for $\kappa$ And}
\subsection{Observations and Basic Data Reduction}
SCExAO targeted $\kappa$ And on UT 8 September 2017  with
 the \textit{CHARIS} integral field spectrograph operating in low-resolution ($R$ $\sim$ 20), broadband (1.13--2.39 $\mu m$) mode \citep{Peters2012,Groff2013}. 
SCExAO/CHARIS data were acquired in pupil tracking/\textit{angular differential imaging} (ADI) mode \citep{Marois2006} with the star's light blocked by the Lyot coronagraph with the 217 mas diameter occulting spot.    Satellite spots, diffractive attenuated copies of the stellar PSF, were generated by applying a 25 nm amplitude modulation on the deformable mirror. \citep{Jovanovic2015b}.   Exposures consisted of 42 co-added 20.6 $s$ frames covering a modest total parallactic angle rotation of  $\sim$ 10.5${\arcdeg}$.    The data were taken under good, ``slow" seeing conditions: 0\farcs{}4--0\farcs{}5 in $V$ band with 2--4 ms$^{-1}$ winds.
   The real-time AO telemetry monitor recorded the residual wavefront error after SCExAO's correction, implying typical exposure-averaged $H$-band 
Strehl Ratios of 90-92\%.  

We used the CHARIS Data Reduction Pipeline \citep[CHARIS DRP;][]{Brandt2017} to convert raw CHARIS data into data cubes consisting 22 image slices spanning wavelengths from 1.1 $\mu m$ to 2.4 $\mu m$.   Calibration data provided a wavelength solution; using the the robust `least squares' method described in \citeauthor{Brandt2017}, we extracted CHARIS data cubes.   Contemporaneous Keck/NIRC2 observations of HD 1160 calibrated CHARIS astrometry, yielding a spaxel scale of  0\farcs{}0162, a $\rho$ $\sim$ 1.05\arcsec{} radius field of view, and north position angle offset of -2.2$^{o}$ (see Appendix A).    

Basic image processing steps -- e.g. image registration, sky subtraction -- were carried out using our CHARIS IDL-based data reduction pipeline, which will later be released alongside a future release of the Python-based CHARIS DRP (i.e. the ``CHARIS Post-Processing Pipeline") and were described in recent SCExAO/CHARIS science/instrumentation studies \citep{Currie2018a,Goebel2018}.    Inspection of the data cubes revealed little residual atmospheric dispersion and exposure-to-exposure motion of the centroid position; the spot modulation amplitude translated into a channel-dependent spot extinction of $atten_{\lambda}$ = 2.72$\times$10$^{-3}$ $\times$($\lambda$/1.55 $\mu m$)$^{-2}$.

To spectrophotometrically calibrate each data cube, we considered both stellar atmosphere models and the widely-used \citet{Pickles1998} library adopted in our previous CHARIS papers \citep{Currie2018a,Goebel2018}, in the GPI Data Reduction Pipeline \citep{Perrin2014}, and in P1640 analysis of $\kappa$ And b's near-IR spectrum in \citet{Hinkley2013}.    As described in the Appendix B, for B9V and some other spectral types the \citeauthor{Pickles1998} library lacks direct measurements in the near-IR and instead adopts an extrapolation from shorter wavelengths that would translate into a miscalibrated companion spectrum.    As an alternative, we used a Kurucz stellar model atmosphere \citep{CastelliKurucz1993}.   Parameters were tuned to closely match those determined from interferometry \citep{Jones2016}: $T_{\rm eff}$ = 11000 $K$, log(g) =  4.0\footnote{The nearest \citeauthor{Pickles1998} model with complete near-IR coverage (A0V) or Kurucz models at slightly different temperatures/gravities (e.g. $T_{\rm eff}$. = 10500, log(g) = 4.5) yielded an identical calibration to within $\sim$ 2\% across the CHARIS bandpass.}. 

\begin{figure}
\centering
\includegraphics[scale=0.325,clip]{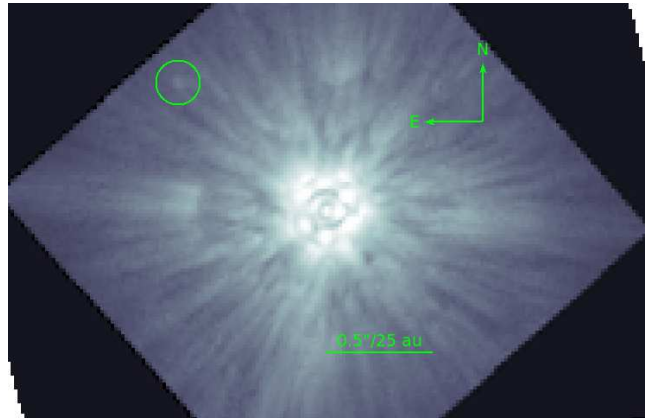}
\caption{A characteristic broadband (wavelength-collapsed) CHARIS image shown in a log color stretch (minimum value to maximum value).   The companion $\kappa$ And b is visible without any PSF subtraction techniques or even unsharp masking applied.   The stellar halo is well suppressed at an intensity roughly or just slightly higher than that of $\kappa$ And b down to $\rho$ $\approx$ 0\farcs{}5.}
\label{kapandraw}
\end{figure}

As shown in Figure \ref{kapandraw}, $\kappa$ And b is visible in raw CHARIS data, with a peak emission roughly three times (0.5--5 times) that of the local speckle intensity in wavelength-collapsed images (individual channels).   In $H$ band, the companion is about as well separated from the speckle halo as it was in earlier, Fall 2016 SCExAO/HiCIAO data obtained with the vortex coronagraph shown in Figure 6a of \citet{Kuhn2018}.   Inspection of our raw broadband images shows that $\kappa$ And b would be marginally visible without processing at smaller separations down to $\rho$ $\sim$ 0\farcs{5}.

\subsection{Point-Spread Function (PSF) Subtraction and Spectral Extraction}
\begin{figure*}[ht!]
\centering
\includegraphics[scale=0.35,clip,trim=15mm 0mm 15mm 0mm]{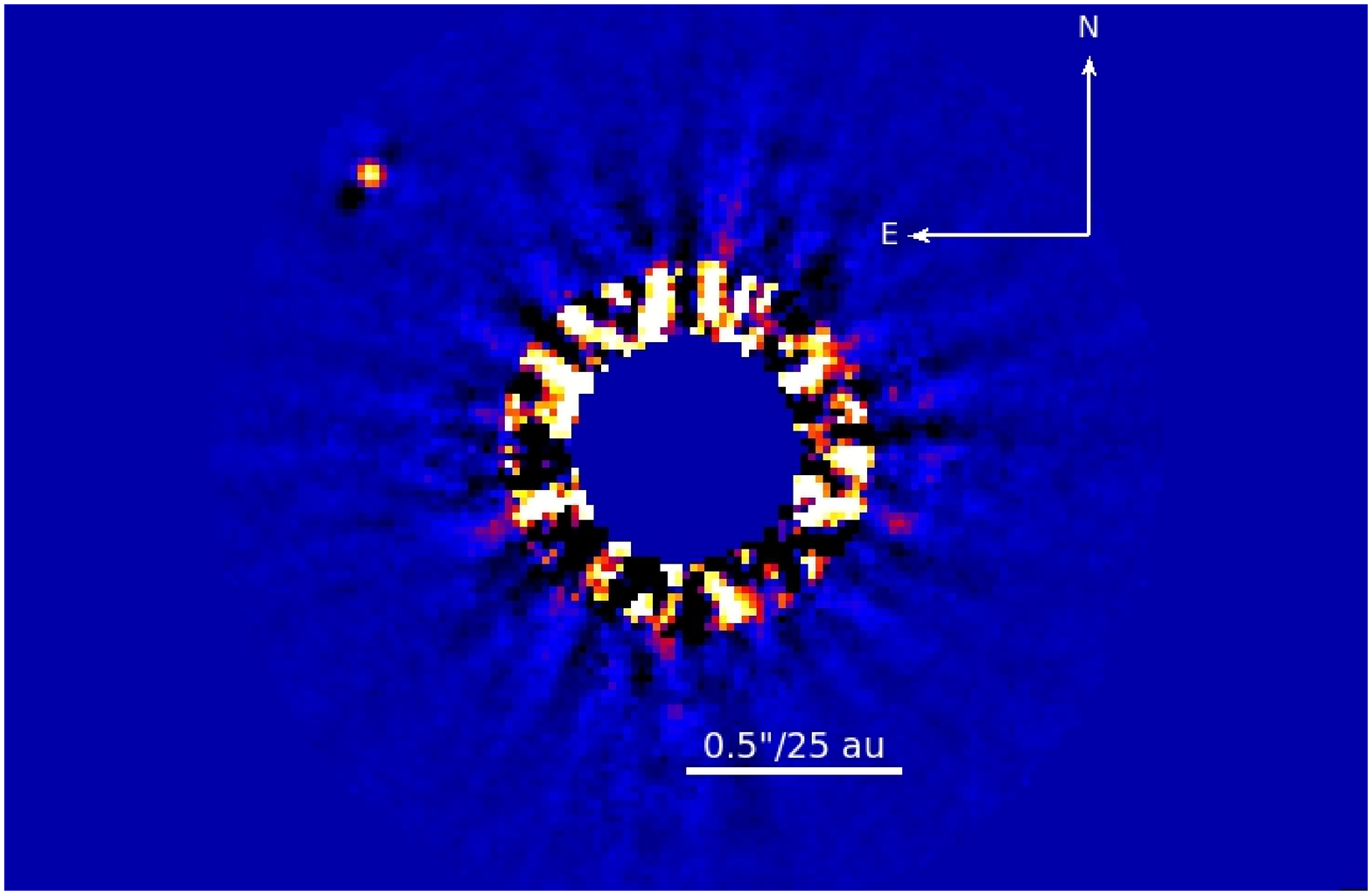}
\includegraphics[scale=0.175,clip,trim=30mm 0mm 30mm 0mm]{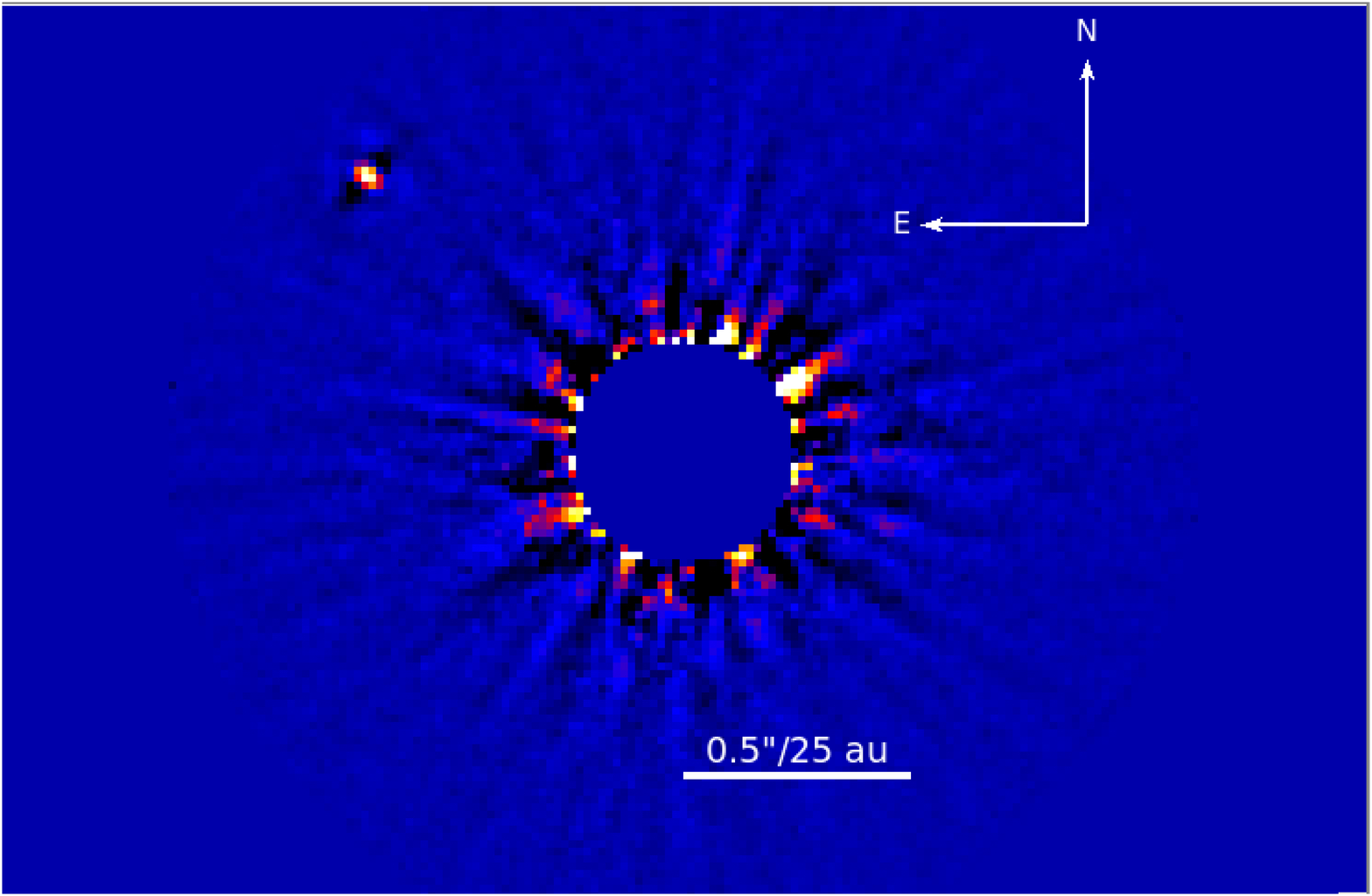}
\caption{Detection of $\kappa$ And b from SCExAO/CHARIS utilizing only ADI (not SDI) (left) with very conservative settings in A-LOCI for PSF subtraction and (right) with a more aggressive reduction using A-LOCI.   $\kappa$ And b is detected with a SNR of 88 and 110.   We extract the spectrum of $\kappa$ And b from the conservative reduction.   The residuals are significantly higher at $\rho$ $\sim$ 0\farcs{}3--0\farcs{}5 for the conservative reduction largely due to a combination of local masking, aggressive covariance matrix truncation, and the large rotation gap.   Throughput is lower and self-subtraction footprints along the azimuthal direction are stronger for the aggressive reduction due to its lack of masking, its less aggressive covariance matrix truncation, and smaller rotation gap.}
\label{kapanddetadi}
\end{figure*}
To further suppress the stellar halo
 and yield a high signal-to-noise (SNR) detection of $\kappa$ And b in each channel,
 we employed advanced point-spread function (PSF) subtraction techniques.    We performed PSF subtraction using \textit{Adaptive, Locally-Optimized Combination of Images} \citep[A-LOCI][]{Currie2012} -- a derivative of the LOCI algorithm \citep{Lafreniere2007a}\footnote{We did not use the Karhunen-Lo\'eve Image Projection (KLIP) algorithm \citep{Soummer2012}.   At full rank (i.e. directly inverting the full covariance matrix), although (A-)LOCI and KLIP use different formalisms they are mathematically equivalent; using SVD to compute the pseudo-inverse of the covariance matrix in (A-)LOCI is similar to truncating the basis set in KLIP \citep{Marois2010b,Currie2014b,Currie2014c,Savransky2015}.   In previous direct comparisons, A-LOCI tended to yield higher SNR detections (up to a factor of 2--3) \citep[e.g.][]{Rameau2013} and more whitened residual noise.    However, in practice, the algorithms simply differ in setup: in whether they use optimization/training zones to construct a PSF model removed from a smaller subtraction zone region, perform masking, and/or use correlation-based frame selection.}.    In this approach,  the PSF $I$ of image slice $i$ in an annular region $s$ is subtracted from a weighted linear combination of other image slice regions  $\{j\}$ in the sequence:
 \begin{equation}
{\cal R}_{i,s} = I_{i,s} - \sum_j \alpha_{ij,s} I_{j,s}~.
\label{eq:loci_sub}
\end{equation}
In LOCI, the coefficients $\alpha_{ij,s}$  are determined solving from a system of linear equations that minimize the residuals between the target slice and references in an ``optimization" region $o$, the solution to the linear
system
\begin{equation}
\mathbf{A} \cdot \boldsymbol{\alpha} = \mathbf{b}~,
\label{eq:axb}
\end{equation}
where the covariance matrix \textbf{A}  and column matrix \textbf{b} are
\begin{equation}
{\rm A}_{jl} = \sum_{{\rm pixels}~k} I_{jk,o} I_{lk,o} \quad {\rm and} \quad
{\rm b}_j = \sum_{{\rm pixels}~k} I_{ik,o} I_{jk,o}~,
\label{eq:axbarrel}
\end{equation}
by a simple matrix inversion.
  The subtraction zone  $s$ is typically a subset of pixels comprising optimization region $o$.  The set of image slices $J$ used to construct a weighted reference PSF is typically defined by those fulfilling a rotational gap criterion, where a point-source in region $s$ has moved some fraction of a PSF footprint, $\delta$$\times$$\theta_{FWHM}$, between frames $i$ and $j$ due to parallactic angle motion.
  
In A-LOCI, this approach is modified in several ways.   First, it optionally 
removes pixels within the subtraction zone $s$ from the optimization zone $o$, which increase point source throughput and, as shown in Appendix C makes algorithm forward-modeling more tractable
 \citep[``local masking"/``a moving pixel mask";][]{Marois2010b,Currie2012}.   Second, it redefines the covariance matrix \textbf{A} and column matrix \textbf{b}, selecting the $n$ image slices best-correlated with the target image slice over each region $o$.   Third,  it rewrites \textbf{A} using \textit{singular value decomposition} (SVD) as \textbf{U}\textbf{$\Sigma$}\textbf{V},  truncating the diagonal matrix , $\Sigma$, at singular values greater than some fraction of the maximum singular value ($svd_{lim}$) before inverting and thus allowing a low(er)-rank approximation of the covariance matrix \textbf{A}:
 \begin{equation}
  \boldsymbol{\alpha} = (\mathbf{U}\mathbf{\Sigma_{> \mathit{svd_{lim}}}}\mathbf{V})^{-1}\cdot\mathbf{b}. 
  \label{eq:invert}
  \end{equation}
   
    We performed two reductions: 1) a conservative one focused on obtaining a high-fidelity spectrum and 2) an aggressive one that maximizes the achieved contrast in our data.  
 In our first ``conservative" approach,  
 we processed data in annular regions for each wavelength channel independently (ADI-only).   The annular subtraction zone of depth $dr$ = 10 was masked,  a weighted reference PSF was constructed from a 75 PSF footprint ``optimization" area exterior to the subtraction zone, and the diagonal terms of the covariance matrix were truncated at  $svd_{lim}$ = 2$\times$10$^{-6}$$\times$max($\Sigma$).  
  In our second ``aggressive" approach, we performed an A-LOCI reduction first utilizing ADI only and then performing \textit{spectral differential imaging} (SDI) on the ADI residuals.   
  For the ADI component, we shrunk the rotation gap, optimization area, and SVD cutoff, leaving
   the $dr$ = 5 pixel-deep subtraction zone unmasked. 
    For the SDI component, we scaled each image slice in the ADI-reduced data cube by wavelength and subtracted the residuals with A-LOCI.  Instead of an angular gap, we imposed a radial gap of $\delta$ = 0.65, masked the subtraction zone, and constructed a weighted reference PSF from pixels at the same separation as the subtraction zone but different angles as in \citet{Currie2017a} from the pseudo-inverse of \textbf{A} truncated at $svd_{lim}$ = 1$\times$10$^{-6}$$\times$max($\Sigma$).   In all cases, given the limited number of exposures, we did not truncate the reference set by cross-correlation.  Finally, we de-scaled, rotated and combined the ADI/SDI-subtracted image slices together for a final data cube and final broadband (wavelength-collapsed) image.   

To assess and correct for signal loss of $\kappa$ And b due to processing and thus extract a calibrated spectrum and precise astrometry, we forward-modeled planet spectra through the observing sequence \citep[e.g.][]{Pueyo2016}.    Our formalism extends that of \citet{Brandt2013}, is detailed in Appendix C, and considers both self-subtraction due to displaced copies of the planet signal weighted by coefficients $\alpha_{ij}$ and perturbations of these coefficients $\beta_{ij}$ due to the planet signal.   

\subsection{High Signal-to-Noise Detection of $\kappa$ And b with SCExAO/CHARIS and Extracted Spectrum}
\begin{figure}
\centering
\includegraphics[scale=0.375,clip,trim=15mm 0mm 15mm 0mm]{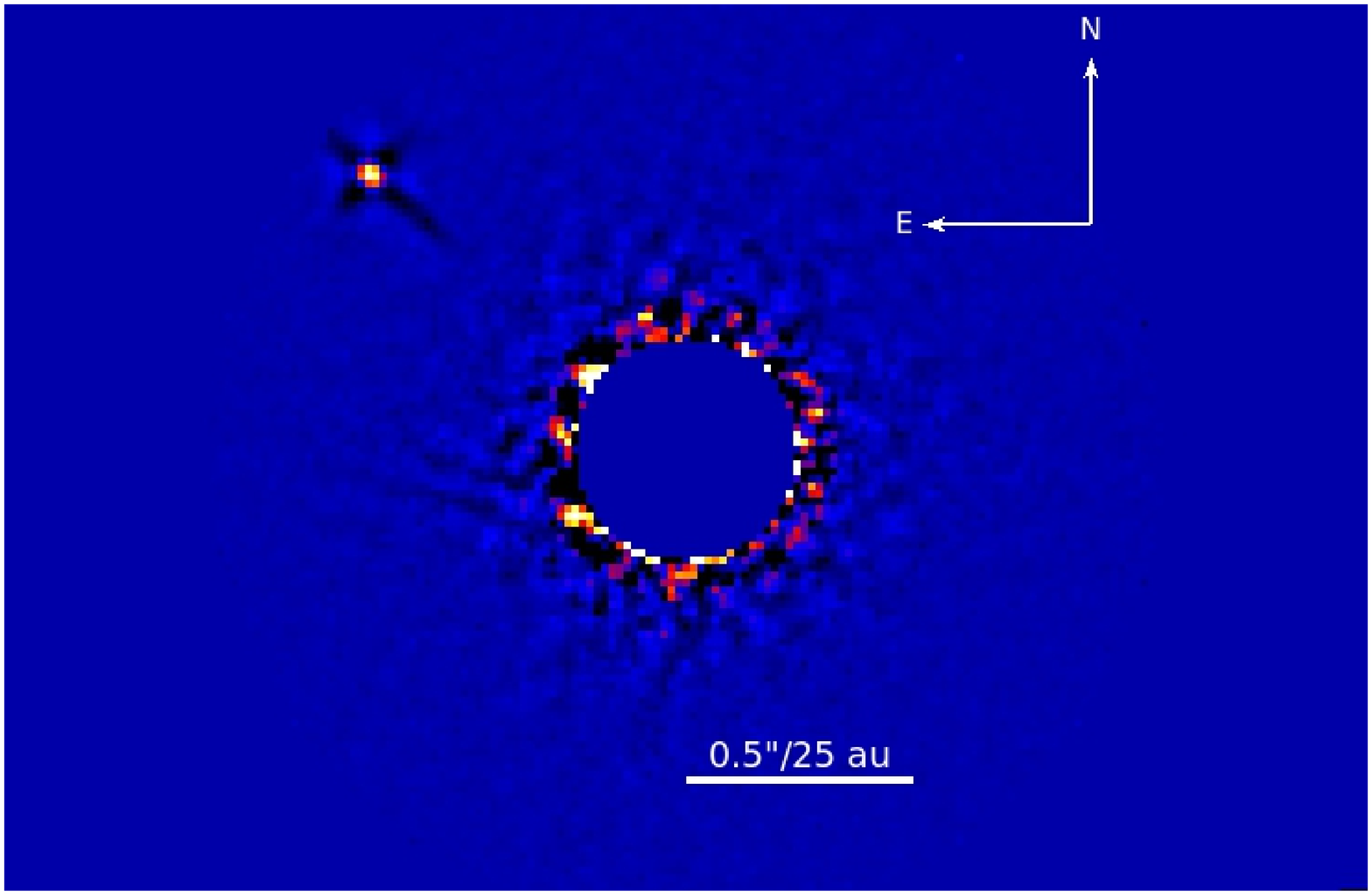}
\caption{Detection of $\kappa$ And b utilizing both ADI and SDI
(SNR $\sim$ 210).   Although the planet now exhibits strong radial self-subtraction footprints due to SDI, its signal loss due to SDI is nearly negligible due to CHARIS' wide spectral bandpass.} 
\label{kapanddetadisdi}
\end{figure}

Figures \ref{kapanddetadi} and \ref{kapanddetadisdi} display wavelength-collapsed CHARIS images reduced using ``conservative" and ``aggressive" PSF subtraction approaches and utilizing ADI only and in conjunction with SDI.  The companion $\kappa$ And b is easily visible at a high SNR (88-210) in the wavelength-collapsed images at a projected separation of $\rho$ $\approx$ 0\farcs{}91 and decisively detected in all channels in all reductions.    Except for channel 6 in the most conservative reduction ($\lambda_{\rm o}$ = 1.376 $\mu$m; SNR $\sim$ 6.4), the detection significance exceeds 10$\sigma$ in all channels for all reductions.   

To extract the spectrum for $\kappa$ And b from the conservative (ADI-only) reduction, we defined the signal from aperture photometry with $r_{\rm ap}$ = 0.5 $\lambda$/D around the best-estimated position (as determined from the wavelength-collapsed image).   We repeated these steps with slight modifications to our algorithm settings to confirm repeatability of the spectrum to a level less than the intrinsic SNR of the detection in each channel.   We confirmed that a negative copy of the extracted planet spectrum, when inserted into our sequence prior to processing, fully nulled $\kappa$ And b in all channels after PSF subtraction.

\begin{figure*}
\centering
\includegraphics[scale=0.5,clip]{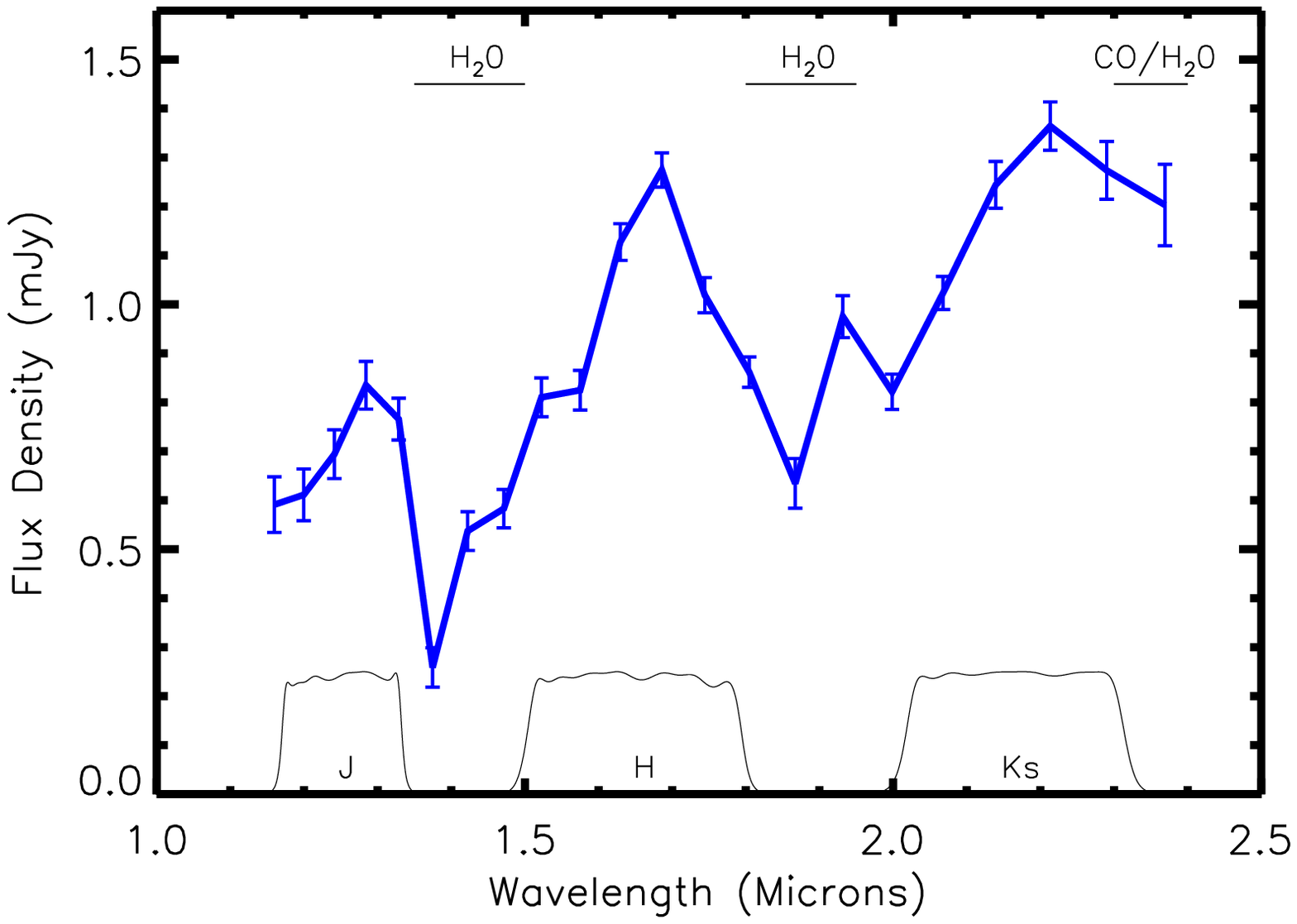}
\includegraphics[scale=0.5,clip]{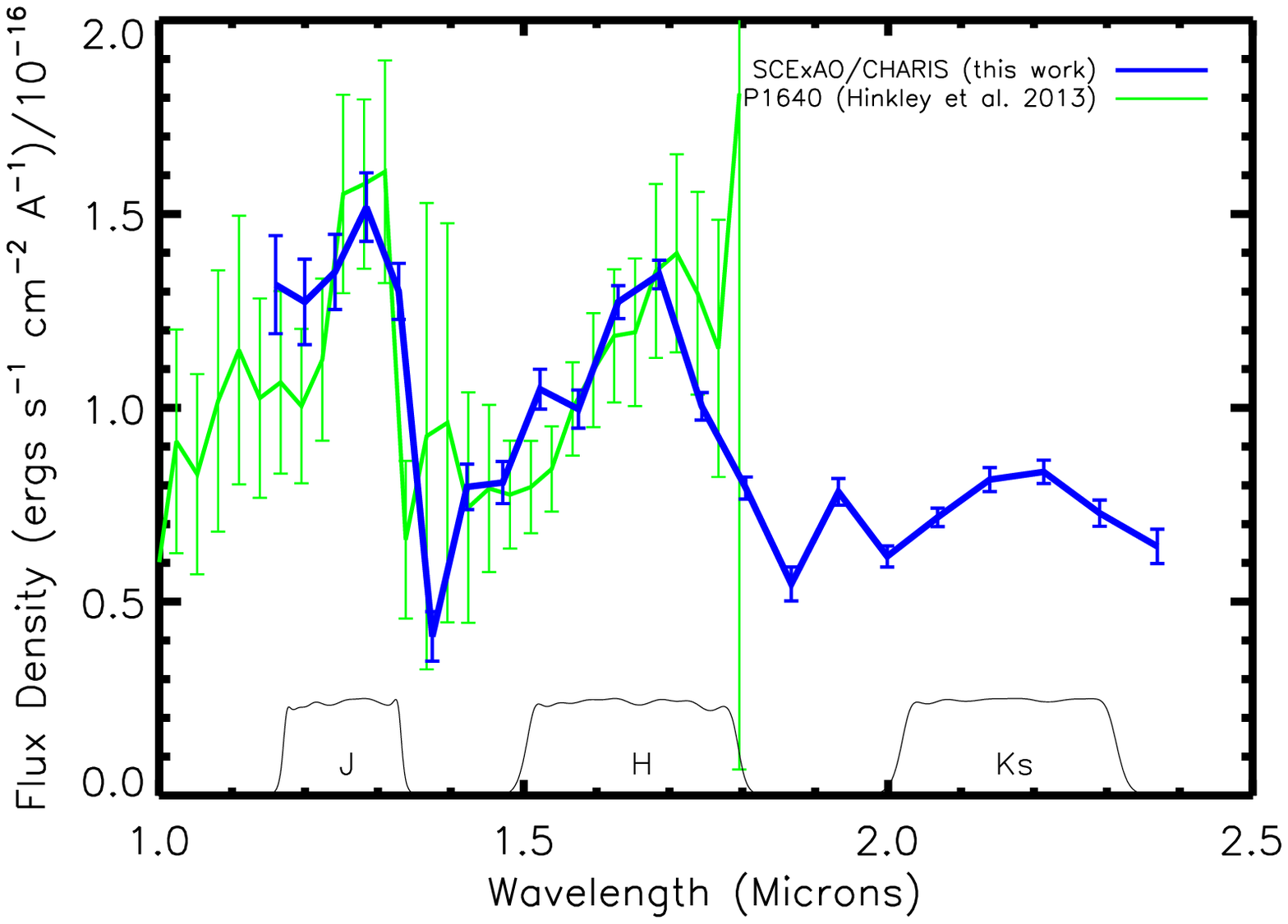}
\caption{(left) SCExAO/CHARIS spectrum of $\kappa$ And b extracted from our conservative (ADI-only) reduction shown in $F_{\rm \nu}$ units with regions attributed major molecular absorption in substellar objects overplotted. (right) SCExAO/CHARIS $\kappa$ And b spectrum compared to that from P1640 presented in \citet{Hinkley2013} and plotted in $F_{\rm \lambda}$ units.  Both panels show transmission profiles for major near-IR passbands $JHK_{\rm s}$ (MKO).   The CHARIS error bars do not include an additional $\sim$ 5\% absolute calibration uncertainty.}
\label{kapandbspectrum}
\end{figure*}

Figure \ref{kapandbspectrum} displays the extracted CHARIS spectrum in units of mJy (left) and (right) compares our spectrum to that from P1640 as extracted in \citet{Hinkley2013} in units of ergs s$^{-1}$ cm$^{2}$ \AA$^{-1}$.   The spectrum is fully listed in Appendix D.   The CHARIS spectrum shows regions of suppressed flux in between the $JHK$ passbands and a slight suppression beyond 2.3 $\mu m$, attributed to water and water/CO absorption in early L dwarfs \citep[e.g.][]{Cushing2005, Cruz2018}.   The $H$ band spectrum is characterized by a clear peak at $\lambda$ $\sim$ 1.65 $\mu$m and steep drop at redder wavelengths; the $K$ band spectrum exhibits a plateau or slightly rising flux between  2.1 and 2.2 $\mu$m.

The CHARIS spectrum shows slight differences with that extracted from P1640 over wavelengths where the two overlap (1.1--1.8 $\mu$m).   The CHARIS spectrum is more peaked in $H$ band than in the P1640 data at $\sim$ 1.65 $\mu$m, with significantly lower flux density at 1.7--1.8 $\mu$m.     Section \ref{Discussion}.1 discusses the sources of these differences.

Following \citet{GrecoBrandt2016}, we assess the nature of residual noise affecting our extracted spectrum by estimating the spectral covariance at $\kappa$ And b's location in our final data cube.   We divided each channel by the residual noise profile and then computed the cross-correlation between pairs of channels $i$ and $j$ in 2 $\lambda$/D-wide annulus at $\kappa$ And b's location, masking pixels within 2 $\lambda$/D of the companion:
\begin{equation}
 \psi_{i,j} = \frac{<\bar{C_{i}{C_{j}}}>}{\sqrt{<\bar{C_{i}^{2}}><\bar{C_{j}^{2}}>}}.
 \end{equation}

\begin{figure}
\centering
\includegraphics[scale=0.65]{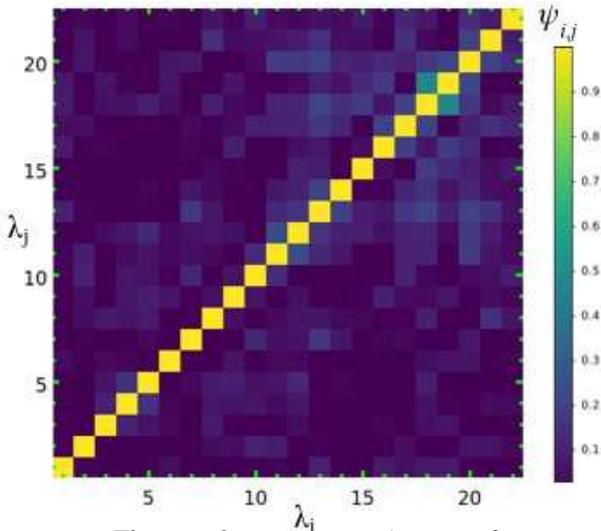}
\vspace{-0.415in}
\caption{The correlation matrix $\psi_{\rm i,j}$ as a function of spectral channel.   Off-diagonal elements identify the effect of residual correlated noise.   With the exception of a few channels (e.g., a slight coupling of channel 17 and 18), the residuals are nearly spatially uncorrelated.}
\label{speccovar}
\end{figure}

Figure \ref{speccovar} displays the spectral covariance at the location of $\kappa$ And b.  Except for a few red channels (e.g. 16 and 17 in K-band), the covariance sharply drops for off-diagonal elements. The functional form for the covariance proposed by \citealt{GrecoBrandt2016} consists of spatially ($\rho$) and spectrally ($\lambda$) correlated noise with characteristic lengths ($\sigma_{\rho}$ and $\sigma_{\lambda}$)  and an uncorrelated term A$_{\delta}$:
\begin{equation}
\psi_{i,j} = A_{\rho}e^{-0.5((\lambda_{i}-\lambda_{j})/\sigma_{\rho})^{2}}+A_{\lambda}e^{-0.5((\lambda_{i}-\lambda_{j})/\sigma_{\lambda})^{2}}+A_{\delta}.
\end{equation}
The data are best fit by A$_{\rho}$ = 0.12, A$_{\lambda}$ = 0.05, A$_{\delta}$= 0.82,  $\sigma_{\rho}$ = 0.65, and $\sigma_{\lambda}$ = 0.24: thus, the residual speckle noise is well-suppressed and poorly coupled between different wavelengths.
At smaller separations where the rotation gap criterion results in far poorer speckle suppression, the noise is dominated by the correlated components (e.g. at $\rho$ $\sim$ 0\farcs{}45, A$_{\rho}$ + A$_{\lambda}$ = 0.56 and A$_{\delta}$= 0.44).

To estimate broadband photometry for $\kappa$ And b, we convolve the spectrum with the Mauna Kea Observatories $JHK$ filter functions binned down to the resolution of CHARIS.
  The companion's apparent magnitude in major MKO passbands is $J$= 15.84 $\pm$ 0.09, $H$ = 15.01 $\pm$ 0.07, and $K_{\rm s}$ = 14.37 $\pm$ 0.07.  Its $J$-$H$ and $J$-$K_{\rm s}$ colors agree with previous estimates from \citet{Carson2013}, \citet{Hinkley2013}, and \citet{Bonnefoy2014a}.   In the 2MASS photometric system, its colors are slightly redder (e.g. $J_{\rm 2MASS}$ - $K_{\rm s, 2MASS}$ = 1.52).

\section{New and Archival Keck/NIRC2 $K_{\rm s}$ Band Astrometric Data}

\begin{figure*}
\centering
\includegraphics[trim=25mm 0mm 25mm 0mm,clip,scale=0.37]{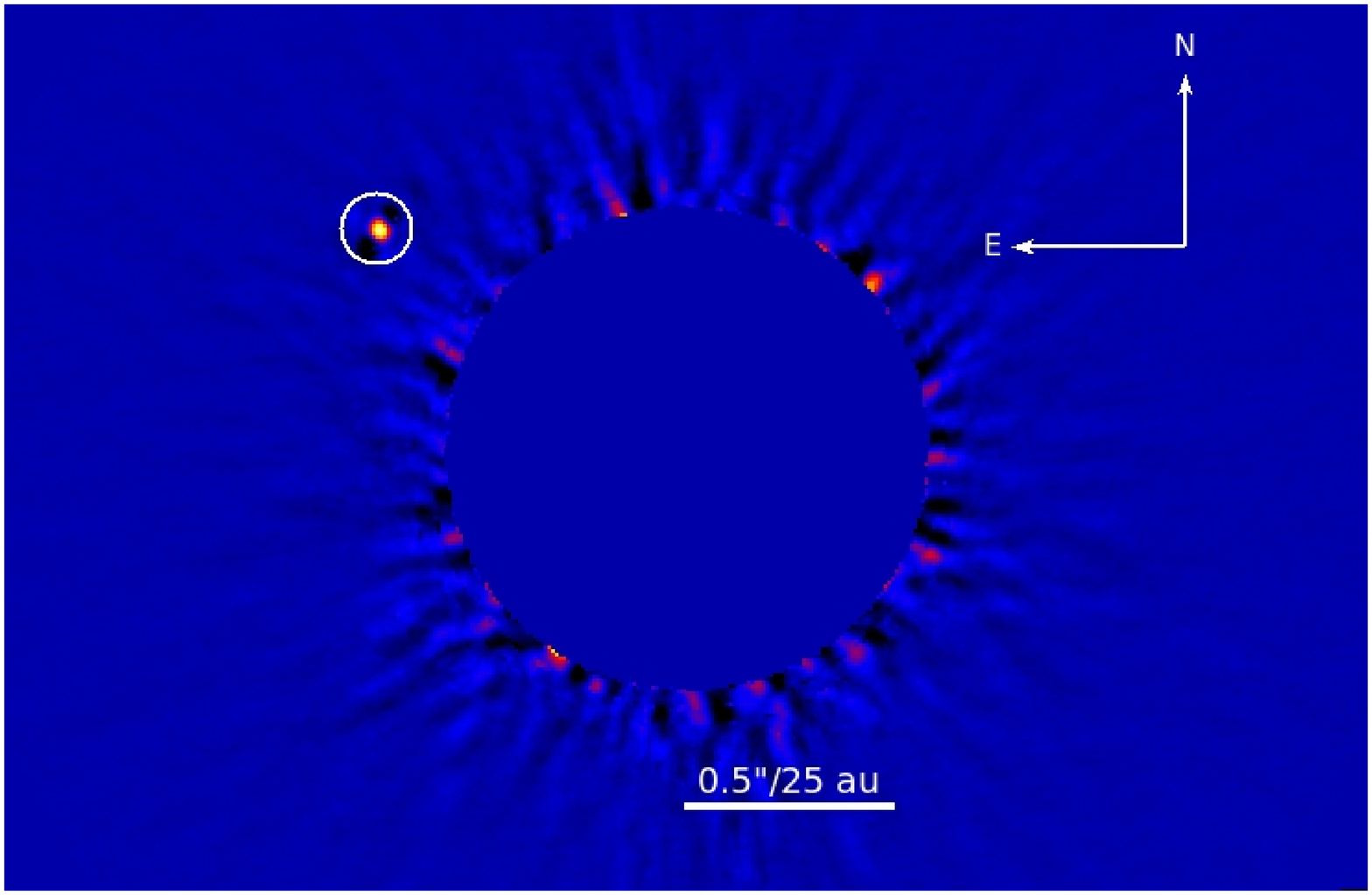}
\includegraphics[trim=25mm 0mm 25mm 0mm,clip,scale=0.37]{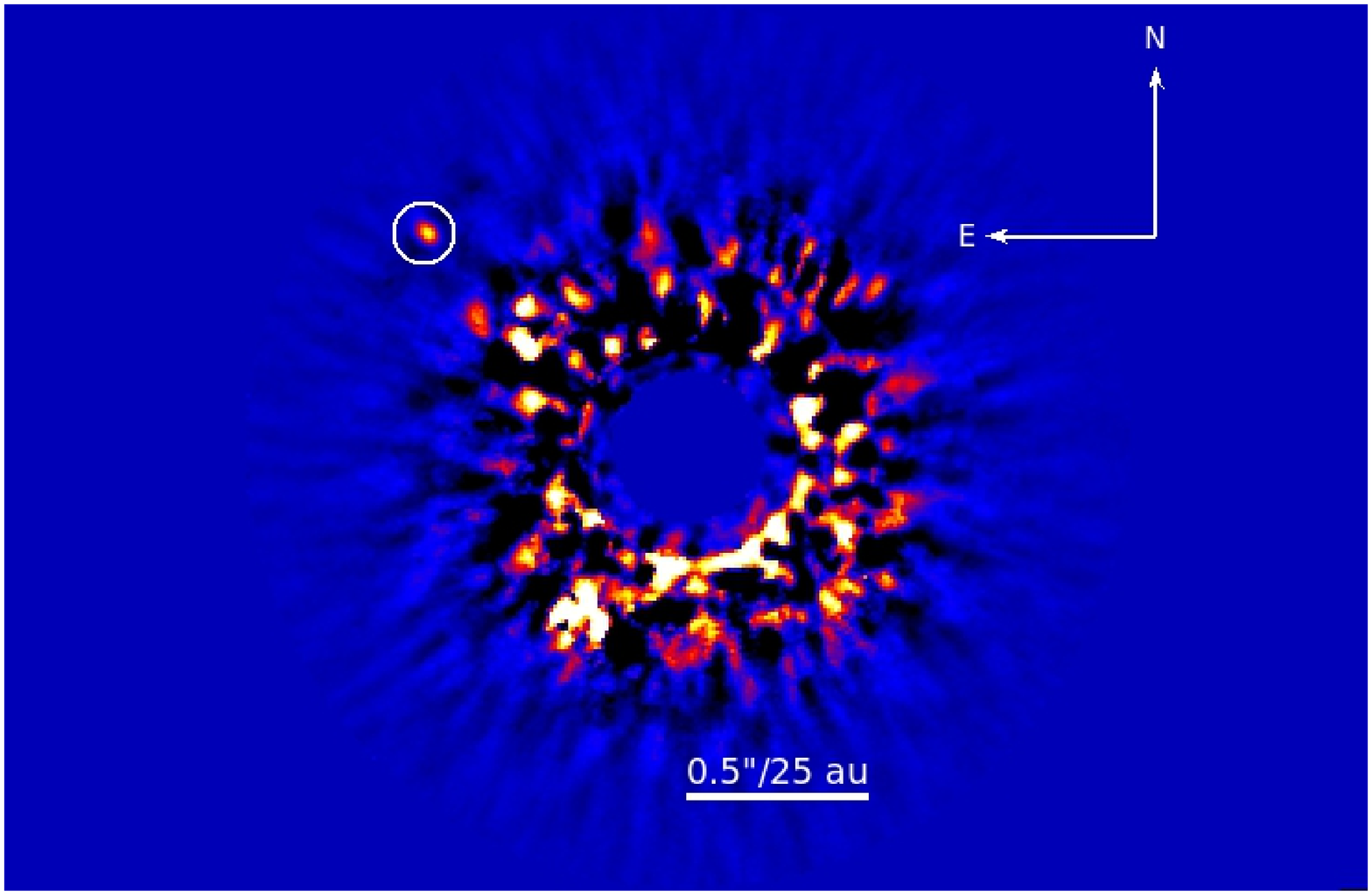}
\caption{(left) Detection of $\kappa$ And b from from archival 2013 Keck/NIRC2 data (PI J. A. Johnson) and (right) from December 2017 Keck/NIRC2 data.    The image scale is equivalent to that in previous figures; $\kappa$ And b is at a wider angular separation in the archival data than in the SCExAO/CHARIS data.}
\label{keckarchive}
\end{figure*}

To supplement $\kappa$ And b's astrometry derived from SCExAO/CHARIS data, we measure its position in well-calibrated data obtained recently and in prior epochs using Keck coupled with the NIRC2 camera.    First, we obtained new Keck/NIRC2 coronagraphic imaging of $\kappa$ And on UT 8 December 2017 in the K$_{\rm s}$ filter using the 0\farcs{}6 diameter coronagraphic spot.   Data consisted of coadded 30-second exposures covering 13.6$^{o}$ of parallactic angle motion.   Basic image processing follows previous methods utilized for $\kappa$ And observations taken with Keck/NIRC2 drawn from \citet{Currie2011a}, including dark subtraction, flat-fielding, distortion corrections, and image registration \citep[][]{Bonnefoy2014a}.   We used A-LOCI with local masking of the subtraction zone to produce a nearly unattenuated detection of $\kappa$ And b (SNR = 27).

Second, we searched for and identified $\kappa$ And $K_{\rm s}$ band data from the Keck Observatory Archive taken on UT 18 August 2013 (PI John Asher Johnson), consisting of 15 20-second exposures.   
A visual inspection of these data reveals $\kappa$ And b, and they fill in the gap in astrometric measurements between the CHARIS data set (September 2017) and those from \citet{Bonnefoy2014a}.
We use A-LOCI with local masking and a rotation gap of 1 PSF footprint to subtract the stellar halo, yielding a high throughput detection and high-precision astrometry.  The SNR is comparable to or slightly higher than that from the discovery paper \citep{Carson2013} and other early detections \citep[e.g.][]{Burress2013}.
Third, we report unpublished astrometry for $\kappa$ And b from data taken on UT 3 November 2012 published in \citet{Bonnefoy2014a}.   

The 2017 and 2013 epoch detections are shown in Figure \ref{keckarchive}.   Astrometry in each data set assumed a 9.971 mas pixel scale and north position angle offset of 0.262$^{o}$ for the 2017 data and 9.952 mas pixel scale and north position angle offset of 0.252$^{o}$ for earlier data sets \citep{Service2016,Yelda2010}.   Comparing the position of $\kappa$ And b in these two data sets and with CHARIS clearly shows that the companion's projected separation is decreasing with time.

\section{Empirical Constraints on $\kappa$ And \lowercase{b}'s Atmospheric Properties}
To analyze $\kappa$ And b's spectrum, we adopt a three-pronged approach, 1) comparing it to optically-anchored L dwarf spectral templates covering a range of gravities, 2) comparing it to a large library of empirical JHK spectra for MLT dwarfs, and 3) assessing gravity from spectral indices.   The templates provide a baseline qualitative assessment for $\kappa$ And b's spectral type and gravity.   The libraries further clarify these parameters, identifying a set of best-fit objects, some of which have well-estimated ages and masses.   The spectral indices serve as a quantitative estimate of gravity.

For empirical comparisons, we quantify the goodness-of-fit by comparing $\kappa$ And b's spectrum $f$ to the $k$th weighted comparison spectrum $F_{k}$, choosing the multiplicative factor $\alpha_{k}$ that minimizes $\chi^{2}_{}$ and considering errors in both $\kappa$ And b and the comparison spectrum:
\begin{equation}
\chi^{2}_{k} = (f-\alpha F_{k})^{T}C_{k}^{-1}(f-\alpha F_{k})
\end{equation}
Here, $C_{k}$ is the covariance matrix, where diagonal terms correspond to measured errors in both $\kappa$ And b ($\sigma_{f}$) and those estimated for the comparison spectrum ($\sigma_{F_{k}}$) and off-diagonal terms consider the coupling of $\kappa$ And b spectral errors between different channels as parameterized in \S 2.3\footnote{The spectrophotometric errors for many library spectra are non-negligible and must be considered calculating the goodness-of-fit.   Similarly, the template spectra from \citet{Cruz2018} are drawn from a collection different sources, and thus the ``template" for a given spectral type should have some uncertainty in each channel.   Thus, the covariance matrix must be recomputed each time a weighted comparison spectrum is fit.}.    

\input{cruz_spectralstandard.tex}

We define acceptly-fitting models as those with a $\chi^{2}$ per degree of freedom less than the 95\% confidence limit: $\chi^{2}_{\nu}$ $\le$ $\chi^{2}_{\nu, 95\% C. L.}$\footnote{\citet{GrecoBrandt2016} discuss the effect of spectral covariance in defining the family of best-fitting solutions quantified by the $\Delta^{2}$ criterion and the 95\% confidence interval about the minimum value and note that the actual $\chi^{2}$ values including covariance can be larger.   As the spectral covariance is low in our case, the diagonal terms dominate and there is only a small difference in $\chi^{2}$ including/not including the covariance.   An analysis adopting a $\Delta^{2}$ instead of $\chi^{2}_{\nu}$ criterion would accept more template and empirical spectra but does not otherwise change our key results about what spectral type $\kappa$ And b best resembles.}   To avoid regions heavily contaminated by tellurics and/or covering wavelengths with missing data, we primarily focused on a set of 16 CHARIS spectral channels covering the MKO JHK bandpasses.    In a second pass, we focus on 11 spectral channels covering $H$ and $K$ only, where broadband spectral features may be diagonistic of gravity \citep{AllersLiu2013, Canty2013}.    In \citet{Chilcote2017}, this is referred to as the ``restricted fit".   Finally, as a check on our results, for empirical comparisons we perform an ``unrestricted" fit the full JHK spectrum, allowing the scaling to freely vary between the three passbands, to account for the intrinsic variation the J--K spectral energy distribution at a given spectral type \citep[e.g.][]{Knapp2004}.

\subsection{Comparisons to Template L dwarf Spectra}
\citet{Cruz2018} compute L dwarf near-infrared spectral average templates, constructed (for each spectral type) from a set from of characteristic optically spectral-typed substellar objects.   The templates cover L0-L4 and L6-L8 field objects, L0-L1 intermediate-gravity dwarfs (L0-L1$\beta$), and low-gravity L0-L4 dwarfs (L0-L4$\gamma$).    The sample of near-infrared spectra comprising each template show typical variations on the order of $\sim$ 5\% across $J$-$K$; inspection of empirical spectra comprising some templates showed variations in spectral shape at similar levels.   Thus, we set a floor to the spectrophotometric uncertainty of 5\%.

  \begin{figure}
  \raggedleft
\includegraphics[trim=27mm 0mm 0mm 0mm,scale=0.41,clip]{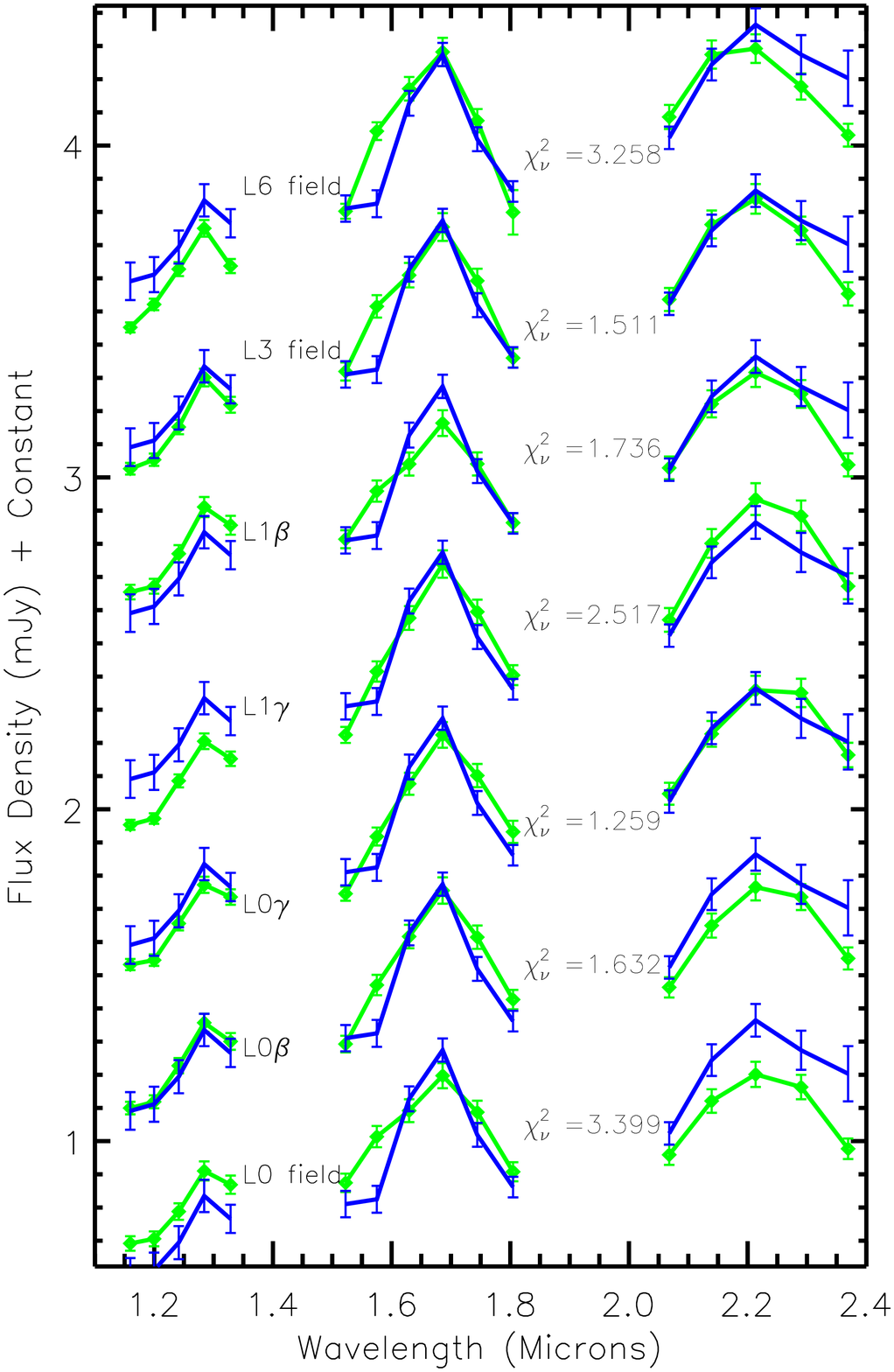}
\vspace{-0.25in}
\caption{
Comparisons between $\kappa$ And b (blue) and spectral templates (green) from \citet{Cruz2018}.   The wavelengths plotted are the sixteen used to define $\chi^{2}$.   The L0$\gamma$ template provides the best fit; L0$\gamma$ and L1$\gamma$ best reproduce the shape of the $H$ and $K$-band portions of the spectra.
 }
\label{templatecompare}
\end{figure}

Table \ref{charisspecstandard} and Figure \ref{templatecompare} compares how well $\kappa$ And b matches each \citeauthor{Cruz2018} template.   Overall, the L0$\gamma$ template best fits $\kappa$ And b's spectrum ($\chi^{2}_{\nu}$ = 1.22), while the L3 field dwarf template marginally fits and the L0--L1$\beta$ templates are marginally inconsistent at the 95\% confidence limit.   When focused more on gravity-sensitive $H$ and $K$ band, low gravity templates  L0$\gamma$ and L1$\gamma$ fit the best;  the L1$\beta$ and L3 field templates are marginally consistent while the L3$\gamma$ and L4 field templates are marginally excluded.     The agreement with the overall shape of $\kappa$ And b's spectrum drives the small $\chi^{2}$ values for the L0$\gamma$ and L3 field templates; the shape of both the $H$ and $K$-band spectra are clearly better fit by the L0$\gamma$ template.

 \begin{figure}
\includegraphics[scale=0.525,trim=11mm 5mm 0mm 0mm,clip]{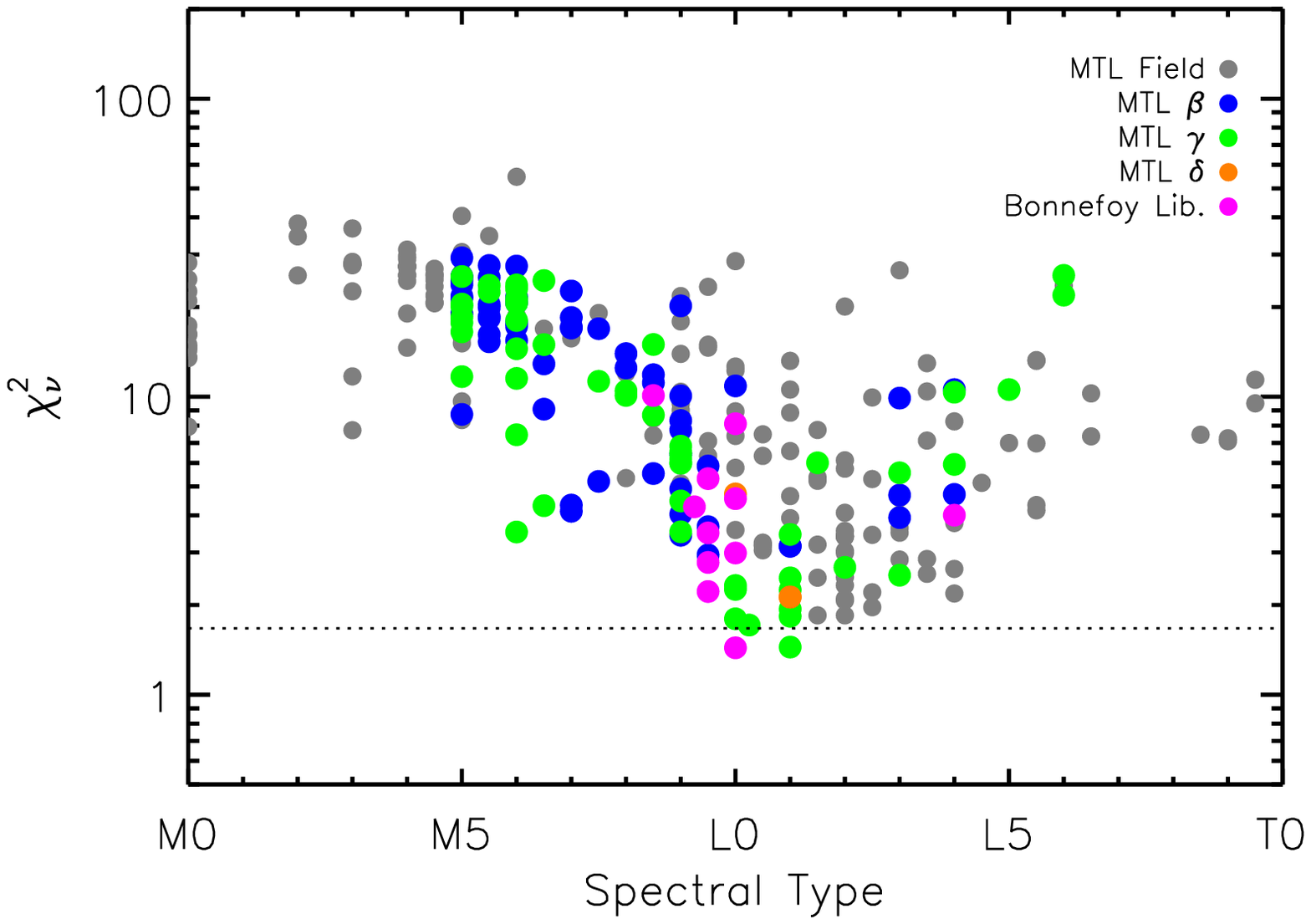}\\
\includegraphics[scale=0.525,trim=11mm 5mm 0mm 0mm,clip]{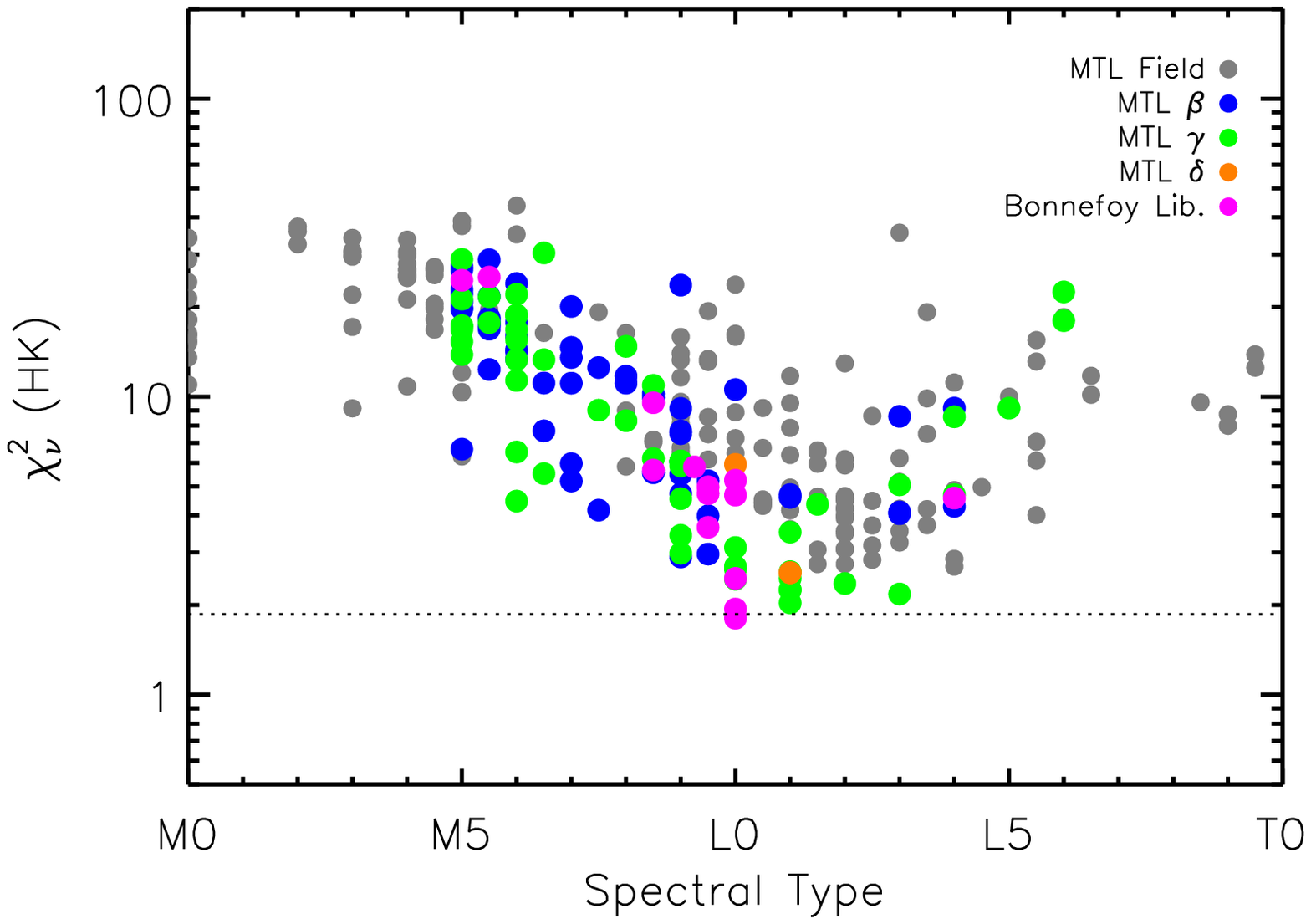}\\
\includegraphics[scale=0.525,trim=11mm 5mm 0mm 0mm,clip]{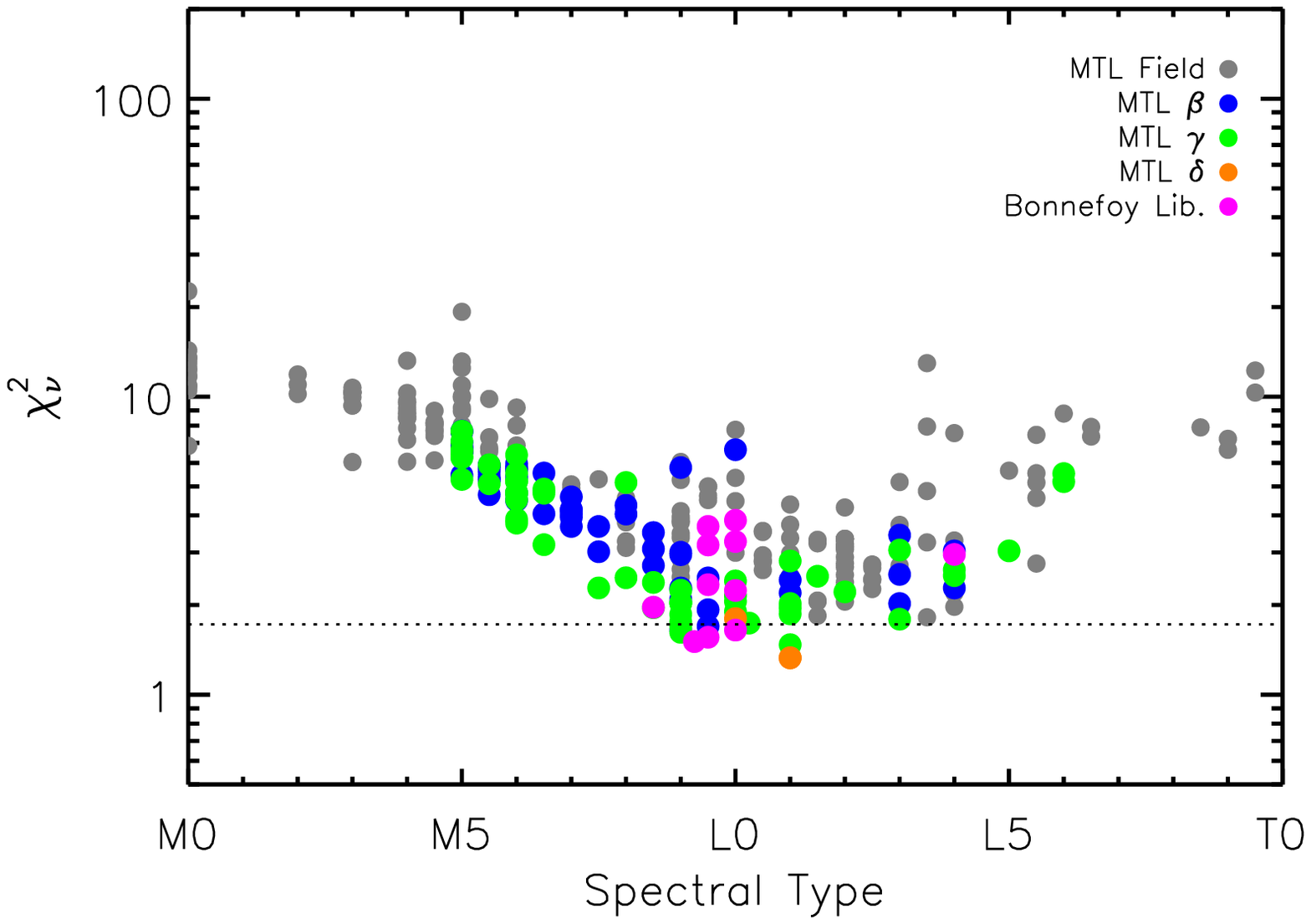}
\vspace{-0.25in}
\caption{(top) For the JHK passbands and just HK (middle), the $\chi^{2}_{\nu}$ statistic comparing $\kappa$ And b to substellar object spectra, including field (gray), intermediate (blue), low (green), and very-low (orange) gravity objects listed in the Montreal Spectral Library and predominantly young, low-mass objects from the VLT/SINFONI library described in \citet{Bonnefoy2014a}.  Horizontal lines identify the $\chi^{2}_{\nu}$ limits below which objects match $\kappa$ And b's spectrum at the 95\% confidence limit.  (bottom) The $\chi^{2}_{\nu}$ distribution when allowing the J, H, and K-band portions of the empirical spectrum to be separately scaled.}
\label{chisqcomp}
\end{figure}

\subsection{Comparisons to Empirical MLT Dwarf Spectra}
Our sample of empirical spectra primarily draws from the Montreal Spectral Library\footnote{\url{https://jgagneastro.wordpress.com/the-montreal-spectral-library/}} and the \citet{Bonnefoy2014b} VLT spectral library
\footnote{\url{http://ipag.obs.ujf-grenoble.fr/~chauving/online_library_Bonnefoy13.tar.gz}}.  
 The Montreal library covers MLT dwarfs with field, intermediate ($\beta$), low ($\gamma$), and very low ($\delta$) gravities characteristic of old ($\sim$ Gyr), intermediate aged ($\sim$ 100 Myr), young ($\sim$ 10--100 Myr), and very young ($<$ 10 Myr) low-mass stars and substellar objects, respectively.   The Montreal data draw from multiple sources presenting spectra reduced using multiple instruments, including \citet{Gagne2014,Gagne2015c}, \citet{Robert2016}, \citet{Artigau2010}, \citet{Delorme2012}, and \citet{Naud2014}.   The Bonnefoy library focuses on objects near the M/L transition (M6-L1) having intermediate to (very-)low gravities ($\beta$$\gamma$$\delta$) with spectra drawn from a single source (VLT/SINFONI) reduced in a uniform manner.    
 We trimmed our Montreal library sample of objects with very low SNR or those with substantial telluric contamination at the edges of the JHK passbands, leaving 360 objects.   Since the \citet{Bonnefoy2014b} library nominally lists a spectrum normalized in $J$ or $HK$, we focused only on those objects whose spectra can be relatively calibrated across $JHK$ (12 objects).

    Figure \ref{chisqcomp} displays $\chi^{2}_{\nu}$ as a function of spectral type for the $JHK$ and $HK$ restricted fits (top and middle panels) and the $JHK$ unrestricted fit (bottom panel), quantitatively showing how well each empirical spectrum matches $\kappa$ And b's spectrum.   The distribution for the restricted fits shows a clear minimum for L0-L1 spectral types with low surface gravity  with 1 (2) objects formally satisfying the 95\% confidence limit for the full $JHK$ ($HK$) spectrum.   In both plots, another 2-3 objects lie just above this limit, all of which are likewise L0-L1 objects with low gravity.   For the unrestricted fit, more objects cluster at or below the 95\% confidence limit, including the $\sim$ 10 $Myr$-old objects UScoCTIO108 B \citep[M9.5 $\gamma$, Bonnefoy library;][]{Bejar2008} and 2MASS J12074836-3900043 \citep[L1$\delta$, Montreal library;][]{Gagne2014}.   The $\chi^{2}_{\nu}$ minimum for the unrestricted fit is broader (M9 to L2--L4), although L0--L1$\gamma$ objects still dominate the subset of those that fit well.  
    
    Table \ref{charisempcompare} lists the best-fitting spectra and their properties from the restricted fits.   
      2MASSJ0141-4633 \citep{Kirkpatrick2006} -- a L0--L1$\gamma$ dwarf and member of the Tucana-Horologium Association-- provides the best fit\footnote{Our adopted spectral type follows estimates from individual indices in \citet{Bonnefoy2014a} rounded to the nearest integer type.}\footnote{Considering the over 500 available spectra in the Montreal library, the L1 dwarf and candidate (10 Myr-old) TW Hya member 2MASSJ1148-2836 numerically provides the best fit to $\kappa$ And b's spectrum.   However, like many other objects, its spectrophotometric errors are very large, and thus it was removed from our model comparisons.}.   In general, the sample of best-fitting objects listed in Table \ref{charisempcompare} is dominated by confirmed and candidate L0--L1$\gamma$ Tuc-Hor members.
    
\begin{figure}
\raggedleft
\includegraphics[trim=20mm 0mm 0mm 0mm,scale=0.315,clip]{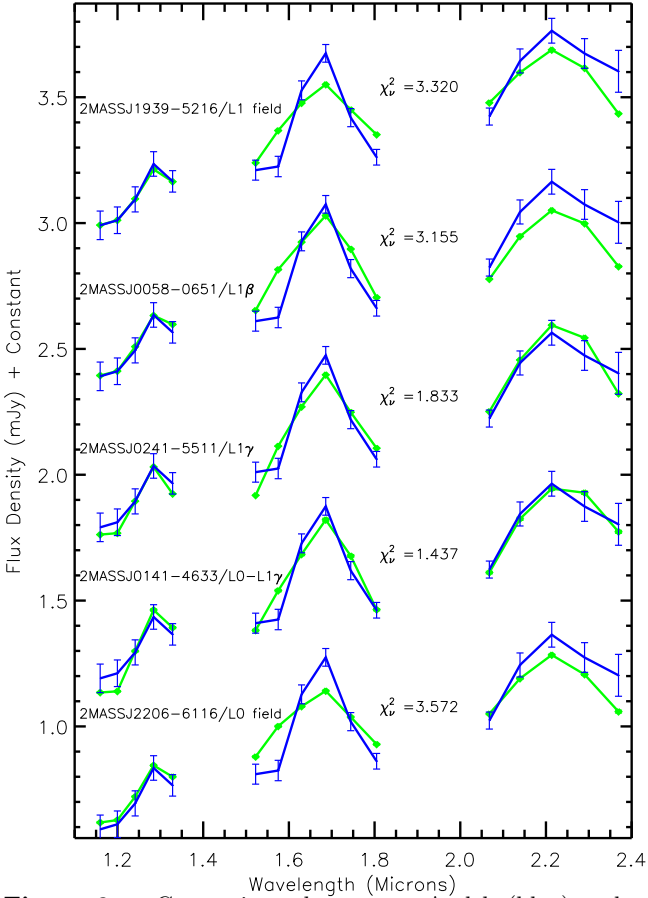}
\vspace{-0.375in}
\caption{
Comparisons between $\kappa$ And b (blue) and a representative sample of L0-L1 dwarfs with different gravity classes (green) from the Montreal and Bonnefoy spectral libraries.    The L0--L1$\gamma$ object 2MASSJ0141-4633 provides the best overall match to $\kappa$ And b. 
}
\label{empcompare}
\end{figure}
    
\input{empcompare_bestfit.tex}

\begin{figure}
\includegraphics[scale=0.45,trim=0mm 10mm 0mm 10mm,clip]{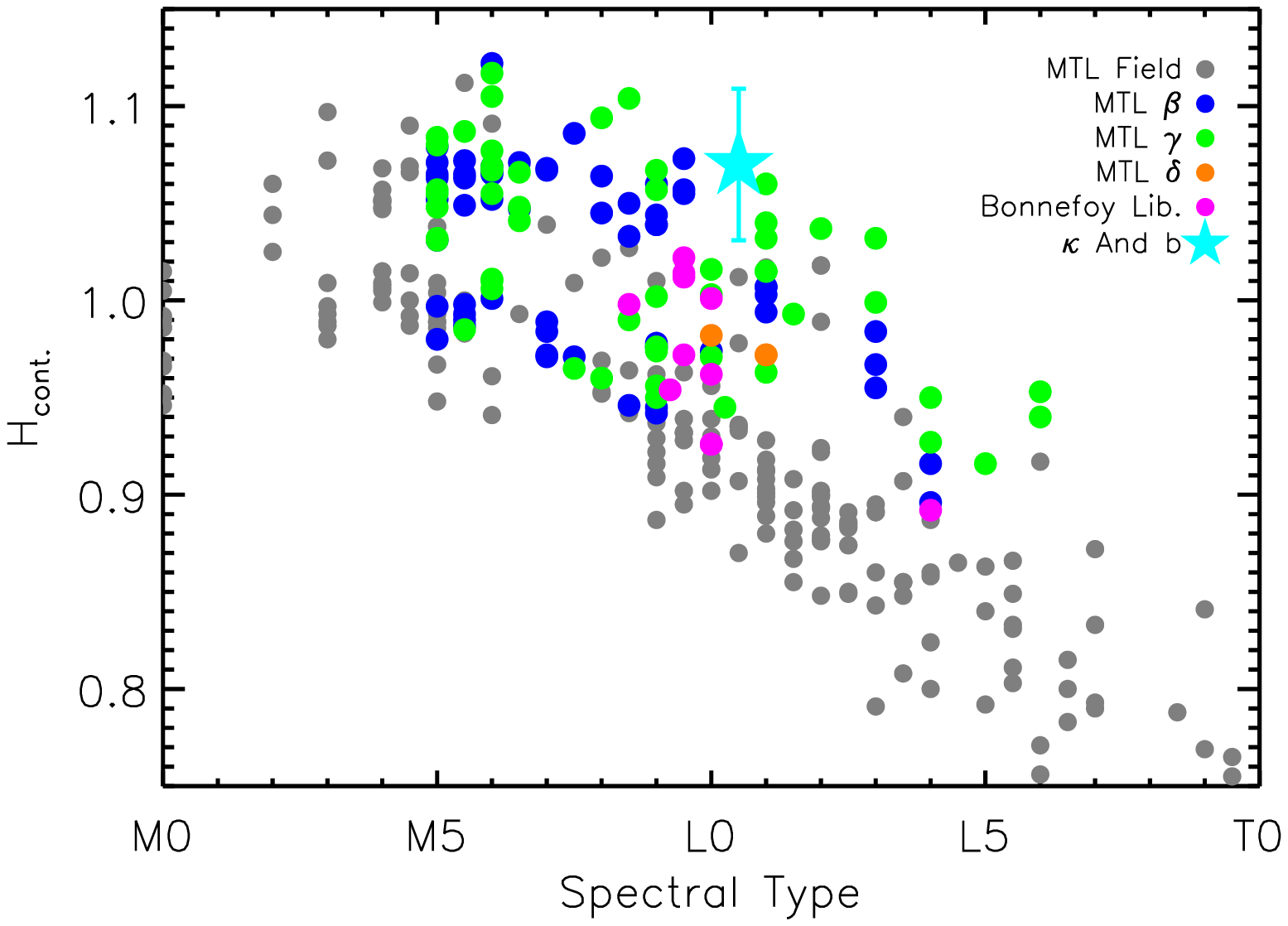}\\
\includegraphics[scale=0.45,trim=0mm 10mm 0mm 10mm,clip]{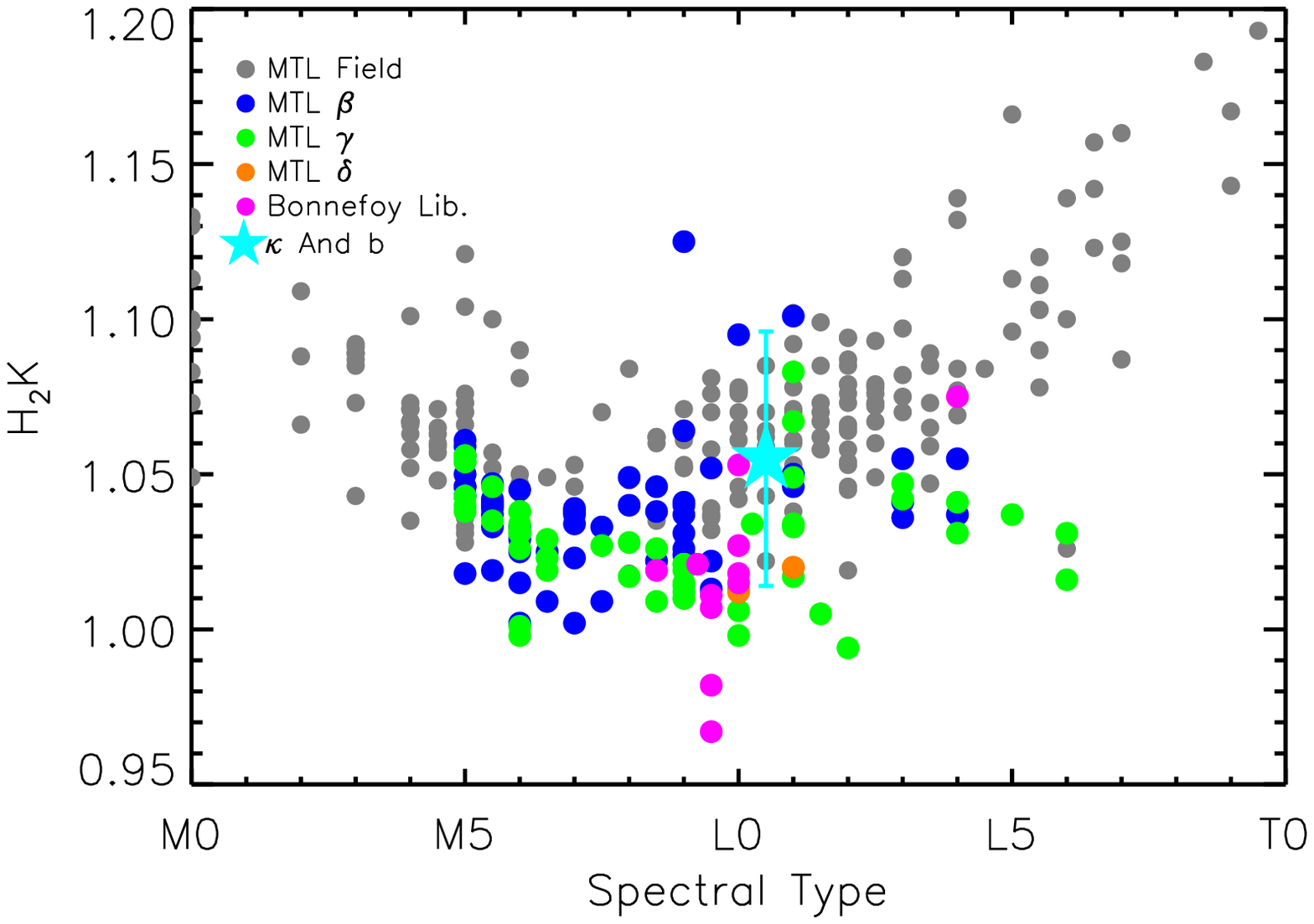}\\
\includegraphics[scale=0.45,trim=0mm 10mm 0mm 10mm,clip]{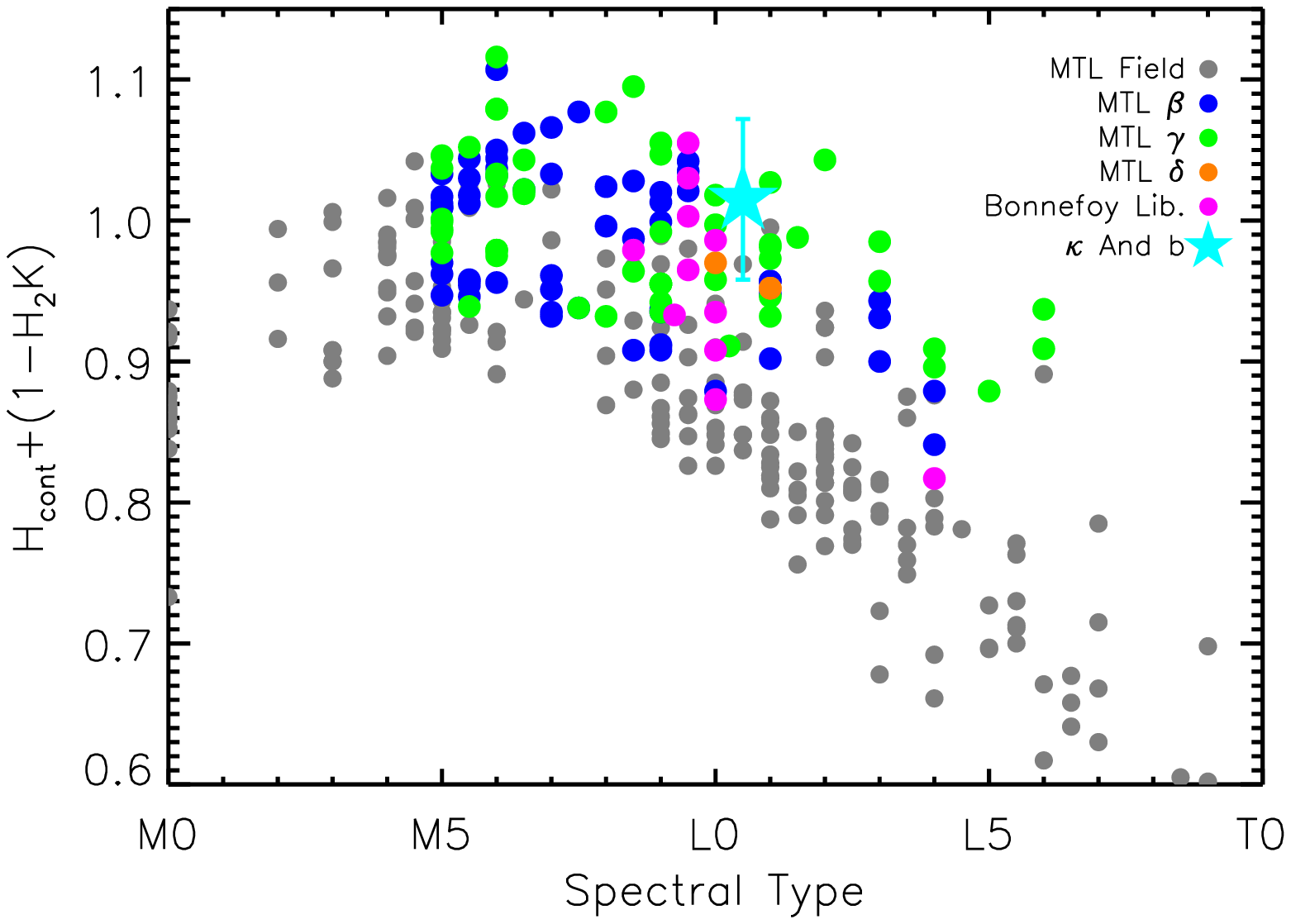}
\vspace{-0.125in}
\caption{Revised (for CHARIS) $H$-continuum (top),  $H_{2}K$ (middle) gravity-sensitive spectral indices, and a combined index (bottom) for $\kappa$ And b (cyan star) and the comparison sample.   Large values for the $H$-continuum   index at a given spectral type suggest low gravity; small values for the $H_{2}K$ index generally low gravity, albeit less decisively.   Uncertainties are shown for $\kappa$ And b; those for the comparison sample are not shown for clarity but are typically on the order of the symbol size.}
\label{gravsens}
\end{figure}

To illustrate how $\kappa$ And b's spectrum best resembles that of a low-gravity L0-L1 dwarf, Figure \ref{empcompare} compares it to 2MASSJ0141-4633 (the best-fitting object with small spectrophotometric errors) and a representative set of L0-L1 objects with small errors and different gravity classes.    
The shape of the $\kappa$ And b $H$ and $K$ spectra strongly favor that of a low-gravity object, as the $H$-band spectrum is far sharper than any field object and the red half of the $K$-band spectrum flatter.   
All other field and intermediate-gravity L0-L1 dwarfs more poorly match $\kappa$ And b.   Other L0-L1$\gamma$ dwarfs have $\chi^{2}$ values that are still characteristically smaller than L0-L1 field objects (see Figure \ref{chisqcomp})\footnote{The major contributor to $\chi^{2}$ for most objects, including the L0--L1$\gamma$ objects displayed, is the $H$-band shape, where $\kappa$ And b has a slightly sharper $H$-band shape.   Some of the youngest, lowest-mass objects better match this feature (e.g. Cha 1109, UScoCTIO 108B) while more poorly matching other parts of the spectrum; a few others (e.g. KPNO Tau 4) have sharper overall H band shapes.}.

\subsection{Quantitative Assessments of Surface Gravity Using Spectral Indices}
We use multiple near-infrared spectral indices to assess the companion's surface gravity:
 the $H$-continuum index ($H$-cont) defined by \citet{Slesnick2004} and the $H_{\rm 2}$$K$ index described by \citet{Canty2013}.   
The $H$-cont index is defined from two measurements of the ``continuum" flux ($\lambda_{1}$ =  1.470 $\mu m$, $\lambda_{2}$ = 1.670 $\mu m$) and a measurement of the ``line" flux at 1.560 $\mu m$:
 \begin{equation}
H_{\rm cont} = \biggl[\frac{\lambda_{\rm line}-\lambda_{\rm 1}}{\lambda_{\rm 2}-\lambda_{\rm 1}}F_{\lambda_{2}}+\frac{\lambda_{\rm 2}-\lambda_{\rm line}}{\lambda_{\rm 2}-\lambda_{\rm 1}}F_{\lambda_{1}}
\biggr]/F_{\rm line}.
 \end{equation}    The $H_{\rm 2}$$K$ index is defined as the flux ratio in two small bandpasses in $K$: H$_{2}$K$_{\rm ind}$ = F$_{\lambda, \rm 2.17 \mu m}$/F$_{\lambda, \rm 2.24 \mu m}$. 

 The wavelengths at which these spectral indices are usually evaluated does not perfectly map onto the wavelengths for each CHARIS channel in low-resolution mode, and the bandpasses width ($\Delta$$\lambda$ $\sim$ 0.02 $\mu m$) is smaller than the change in wavelength between adjacent CHARIS channels ($\Delta$$\lambda$ $\sim$ 0.05 $\mu m$).   Thus, the spectral indices had to be modified.   For $H$-cont, the change is slight: we defined the ``line" flux at channel 10 ($\lambda_{line}$ = 1.575 $\mu m$) and the continuum at channels 8 and 12 ($\lambda_{cont.}$ = 1.471 $\mu m$ and 1.686 $\mu m$).   Wavelengths listed by \citet{Canty2013} for the $H_{2}$K index are more poorly matched to wavelengths defining the CHARIS low-res channels.   We therefore defined an approximate $H_{2}K$ index from averages of adjacent channels 19-20 and 20-21: $H_{2}K$ = ($F_{\lambda = 2.139 \mu m}$ + $F_{\lambda = 2.213 \mu m}$)/($F_{\lambda = 2.213 \mu m}$ + $F_{\lambda = 2.290 \mu m}$).

Figure \ref{gravsens} compares the $H$-cont and $H_{2}$K index for $\kappa$ And b with those from the Montreal and Bonnefoy libraries.      For spectral types of M5 to L6, the typical $H$-cont indices for field dwarfs range from 1 to 0.85.   Indices for young low/intermediate gravity dwarfs from the Montreal and Bonnefoy samples are systematically 0.05--0.10 dex larger, exhibiting very little overlap with the field.  The $H_{2}K$ appears best at selecting very young (t $<$ 10 Myr) objects dominating the Bonnefoy sample \citep[see also][]{Gagne2015c}.   The $H_{2}K$ indices for young low/intermediate gravity dwarfs are less well separated from the field than $H$-cont indices.   However, they are still characteristically smaller than field objects, suggesting that this metric may be used to supplement an assessment of gravity derived from the $H$-cont index.
Combining the two indices together retains a clear separation between nearly all young, low gravity dwarfs and field objects.
   \textit{Thus, although the low resolution of CHARIS broadband mode precludes a direct application of standard metrics for gravity in $H$ and $K$ bands, slightly modified versions of these metrics (especially H-cont) can still identify \underline{likely} young, low-gravity objects}.





The measured gravity-sensitive indices for $\kappa$ And b -- $H$-cont = 1.070 $\pm$ 0.039 and $H_{2}K$ = 1.055 $\pm$ 0.041 -- suggest a low surface gravity.   The $H$-cont index of $\kappa$ And b is larger than any L0-L1 Montreal or Bonnefoy sample object and most similar to L0-L1 objects classified as having a low gravity.   The $H_{2}$K index, which is less diagnostic of surface gravity, is less conclusive since $\kappa$ And b's value overlaps with both field and low gravity objects.   However, considering both indices together, $\kappa$ And b still stands out as an object that best resembles a low-gravity object.

\section{Limits on Additional Companions at Smaller Angular Separations}

\begin{figure}
\centering
\includegraphics[scale=0.5,clip]{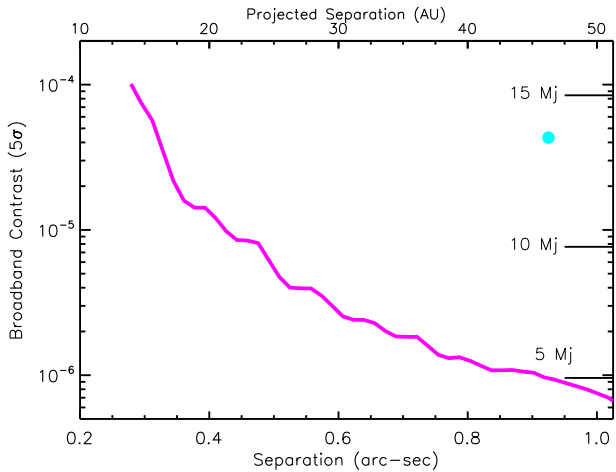}
\caption{The 5-$\sigma$ broadband contrast curve for $\kappa$ And data reduced using ADI+SDI.   A cyan circle identifies the position and contrast of $\kappa$ And b; horizontal bars denote the contrast for substellar objects of various masses.}
\label{contrastcurve}
\end{figure}

Our data do not reveal any additional companions located interior to $\kappa$ And b.   To set limits on companions located interior to $\kappa$ And b, we first divided the 5-$\sigma$ residual noise profile in the wavelength-collapsed ADI+SDI image by the median stellar flux.    We injected model L0 $\gamma$ dwarf spectra from the \citet{Bonnefoy2014a} library and propagated them through ADI and then SDI to determine their signal loss.  We performed ten iterations of forward-modeling and interpolated the results to create a ``flat field" to correct our noise profile map.   Due to CHARIS's large bandpass and our use of local (subtraction zone) masking, signal loss from SDI was minor ($\lesssim$ 20\%), and the radially-averaged throughput ranged between 59\% and 73\% from $\rho$ $\sim$ 0\farcs{}3 to 1\farcs{}0.

To translate our broadband contrast limits to stellar mass, we used the \citet{Baraffe2003} evolutionary models to predict values for gravity and temperature and then atmosphere models to determine the ``broadband" (JHK) flux density for 3--30 $M_{\rm J}$ substellar objects at these gravities/temperatures at 40 $Myr$.   Values ranged from $T_{\rm eff}$ $\sim$ 600 $K$, log(g) = 3.5 to $T_{\rm eff}$ $\sim$ 2300 $K$, log(g) = 4.5.   Atmosphere models draw from A. Burrows, using cloud prescriptions that provide good fits to substellar objects covering most of this range: HR 8799 cde, $\beta$ Pic b, and ROXs 42Bb \citep[][Currie, Burrows et al. 2018 in prep.]{Currie2011a,Madhusudhan2011,Currie2013}.

Figure \ref{contrastcurve} displays our contrast curve.   The broadband contrast dips just below 10$^{-6}$ at wide separations and gradually increases to 10$^{-5}$ at $\rho$ $\sim$ 0\farcs{}35--0\farcs{}45.   Despite extremely poor field rotation and $\sim$ 12 minutes of integration time, our contrasts exterior to 0\farcs{}35--0\farcs{}45 are comparable to those from SCExAO/HiCIAO for HD 36546 -- a factor of 3 deeper and factor of 10 better field rotation \citep{Currie2017a}-- as well as \textit{Gemini Planet Imager} first-light imaging of $\beta$ Pic b, which were likewise much deeper than our data \citep{Macintosh2014}.   Companions with contrasts and masses at or below that of $\kappa$ And b would have been detectable down to 0\farcs{}3 (15 au).   Any companion more massive than $\kappa$ And b and capable of scattering it to wide separations must lie within 15 au.  
   
\begin{figure*}
\centering
\includegraphics[trim=0mm 6mm 65mm 0mm,scale=0.765,clip]{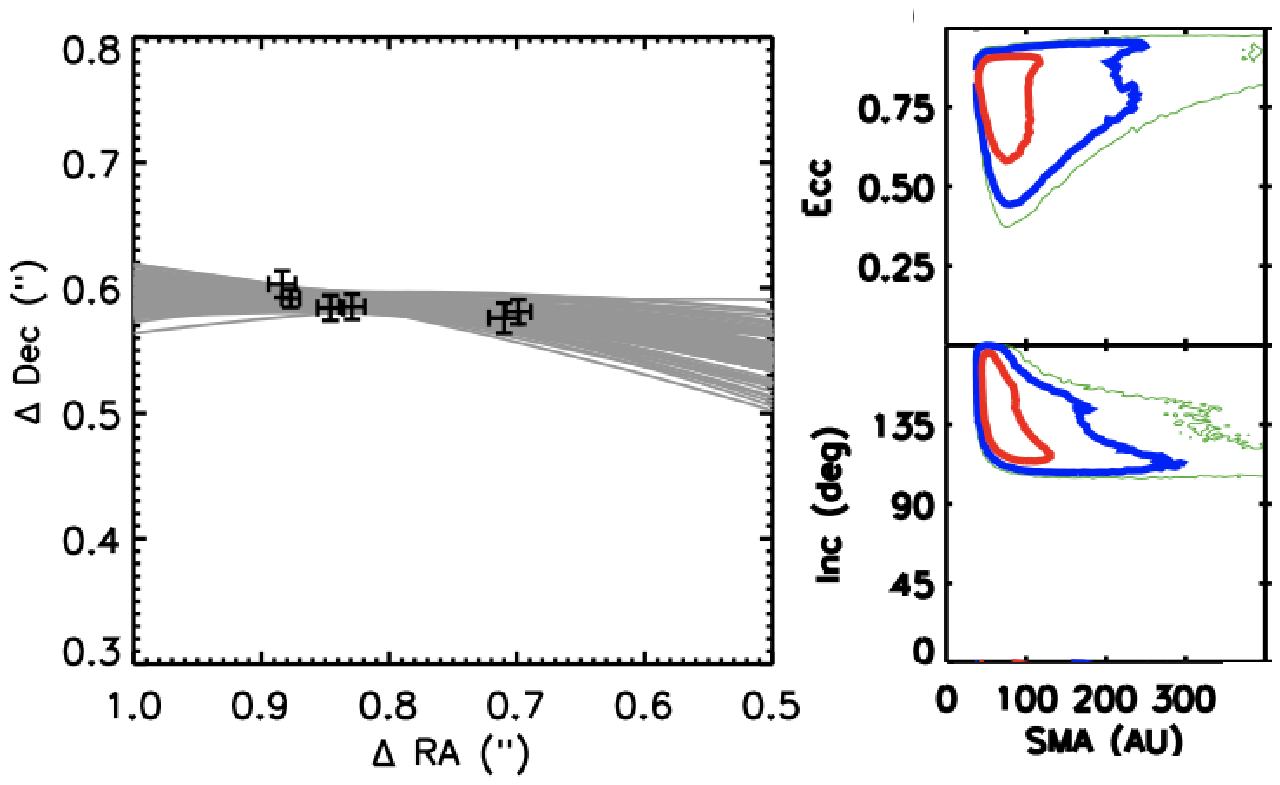}
\includegraphics[trim=0mm 0mm 60mm 0mm,scale=0.365,clip]{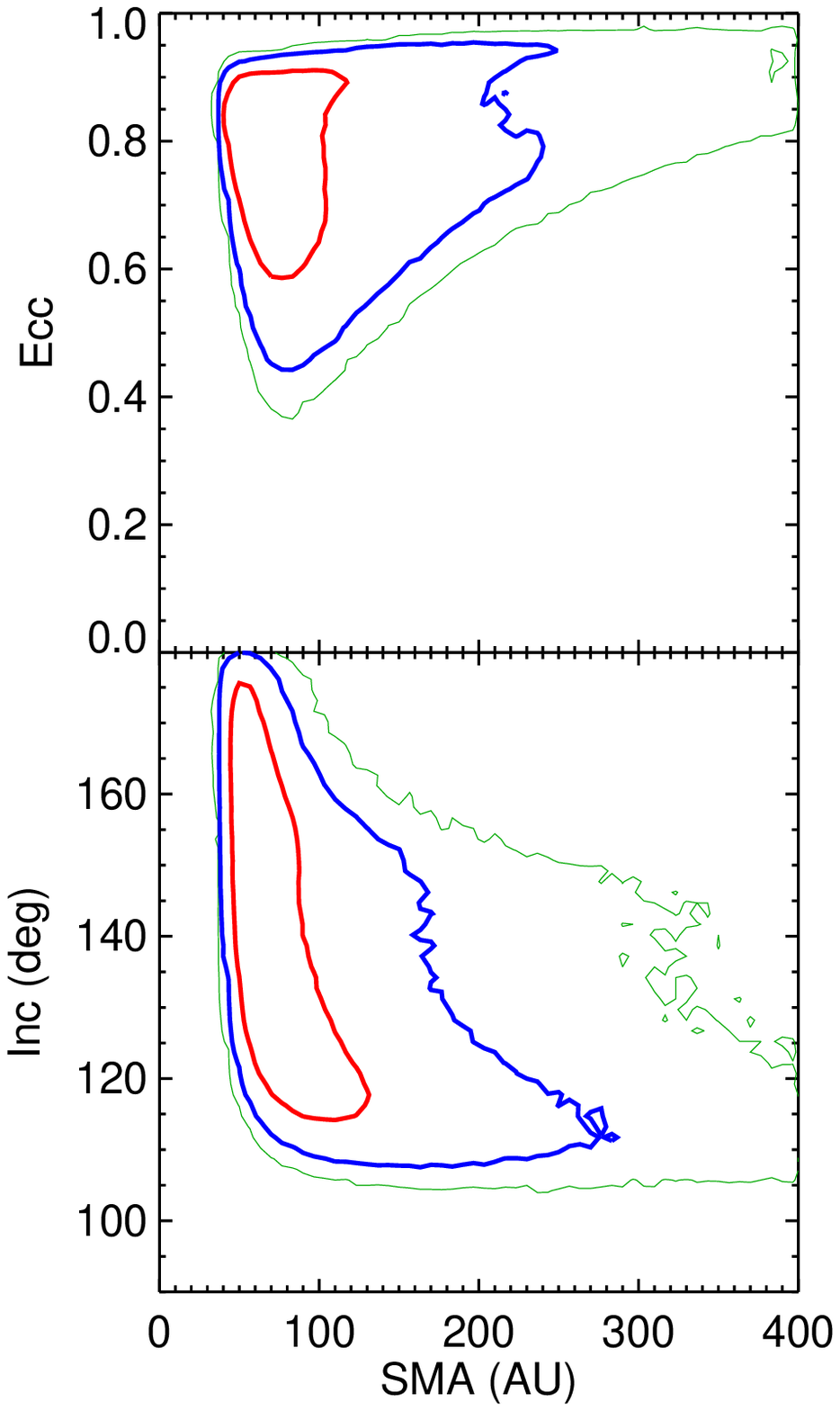}
\includegraphics[trim=0mm 0mm 0mm 0mm,scale=0.3575,clip]{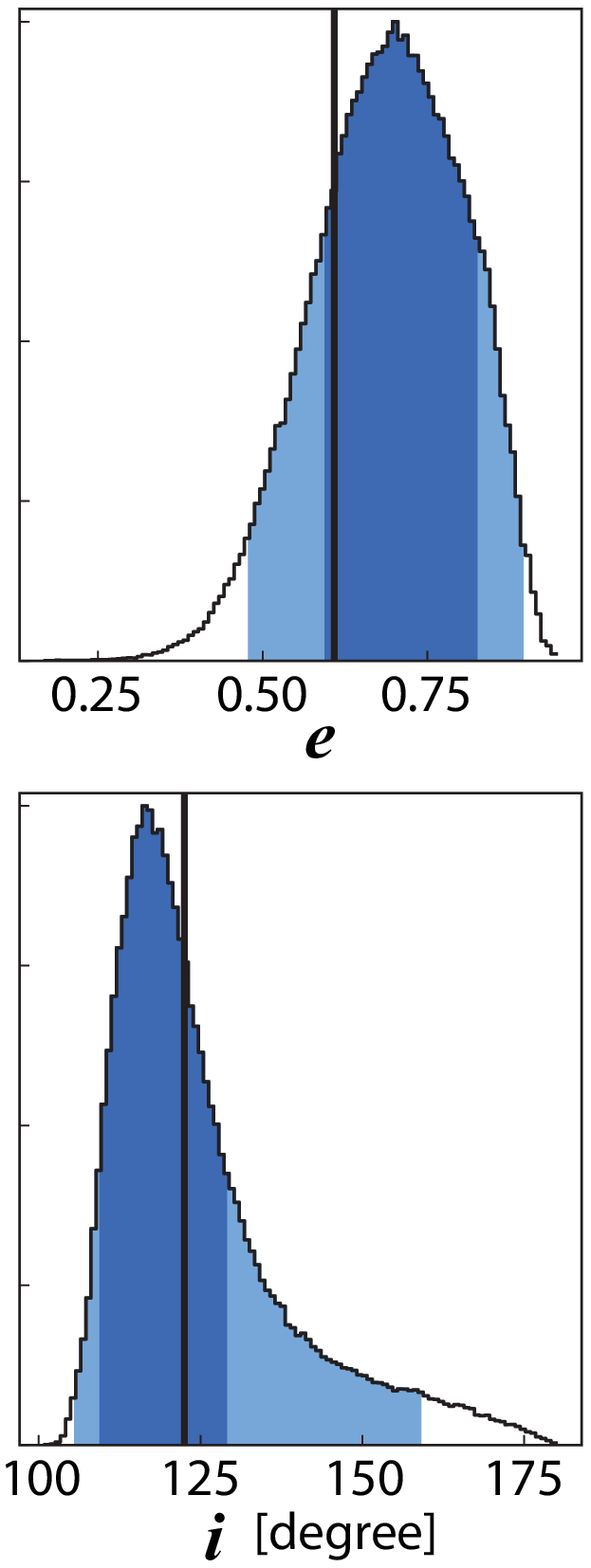}
\vspace{-0.05in}
\caption{Orbit fitting results using OFTI \citep[][]{Blunt2017} and ExoSOFT \citep[][]{MedeBrandt2017}.      (Left) The best-fit orbits from OFTI compared to astrometric data.  The first two epochs draw from HiCIAO astrometry presented in \citet{Carson2013}; the third, fourth, and last epochs are our NIRC2 astrometry while fifth epoch is from CHARIS.  The results and the goodness-of-fits from ExoSOFT are comparable.  (middle) The 68\% (red) and 95\% (blue) confidence intervals for semimajor axis, inclination, and eccentricity from OFTI.  (right) The probability distributions for eccentricity and inclination from ExoSOFT: the dark and light blue regions identify the 68\% and 95\% confidence intervals, respectively.}
\label{orbitresults}
\end{figure*}

\input{orbitfit.tex} 
\section{The Orbit of $\kappa$ And \lowercase{b}}
Well-calibrated astrometry for $\kappa$ And b now spans five years and reveals a clear change in position with time.   Orbital solutions derived for objects with low phase coverage are highly sensitive to input priors on different orbital parameters \citep{KosmoONeil2018}.   We use two different approaches -- OFTI and ExoSOFT \citep{Blunt2017, MedeBrandt2017} -- and adopt different priors to determine plausible orbital properties of the companion.   The first investigation of $\kappa$ And b's orbit was carried out by \citet{Blunt2017}; our focus is to improve upon these constraints using a longer time baseline to determine the companion's orbital direction and identify plausible values for its semimajor axis, eccentricity, and orbital inclination.

OFTI uses a Bayesian rejection sampling algorithm to efficiently determine the most plausible orbital parameters.   We assume Gaussian priors for the parallax centered on \textit{GAIA} DR2 catalogue values, a uniform prior in stellar mass (2.7--2.9 $M_{\odot}$), and impose a log-normal prior in semimajor axis (a$^{-1}$).  ExoSOFT uses a Markov Chain Monte Carlo approach to determine the orbital fit and posterior distributions and \textit{Simulated Annealing} to find reasonable starting positions for the Markov chain and tune step sizes.     We assume a Jeffrey's prior for the semimajor axis (a$^{-1}$/ln(a$_{max}$/a$_{min}$)), which gives equal prior probability for the semimajor axis for each decade of parameter space explored.    Our astrometric errors conservatively consider the intrinsic SNR of the detection, uncertainties in image registration, uncertainties due to self-subtraction/annealing, and absolute astrometric calibration.  

Figure \ref{orbitresults} shows orbital fits using OFTI and ExoSOFT and Table \ref{orbitfits} lists the median value for orbital parameters and their 68\% confidence intervals.   Both approaches determine that $\kappa$ And b orbits clockwise on the plane of the sky, likely has a semimajor axis substantially larger than its projected separation (e.g. 76.5$^{o}$$^{+51.7^{o}}_{-19.8^{o}}$ at the 68\% confidence interval for OFTI) and is highly eccentric (e.g. $e$ $\sim$ 0.69$^{+0.14}_{-0.10}$ for ExoSOFT), although astrometric offsets from different epochs can in principle mimic a non-zero eccentricity.   OFTI finds a wide range of acceptable orbital inclinations -- 119.6$^{o}$--157.4$^{o}$ (111.1$^{o}$--171.5$^{o}$ within the 95\% confidence interval)  -- meaning that $\kappa$ And b's orbit is likely inclined $\sim$ 30--70$^{o}$ from face-on:  a subset of these solutions could imply that the companion's orbital plane is aligned with that of the star ($i_{eq}$ $\sim$ 60$^{o}$).    ExoSOFT finds slightly lower inclinations although orbits aligned with the star's rotational axis lie within the 95\% confidence interval.   

\section{Discussion}\label{Discussion}
\input{kapand_prop.tex}
\subsection{New Constraints on the Atmosphere and Orbit of $\kappa$ And b}
Our study clarifies the atmospheric and orbital properties of $\kappa$ And b, summarized in Table \ref{kapandprop}.   Previous studies analyzing broadband photometry and P1640 spectra \citep{Bonnefoy2014a,Hinkley2013} admit a wide range of acceptable spectral types or different answers depending on a) whether field or low/intermediate gravity comparison spectra are used or b) the wavelength range used for matches with empirical spectra\footnote{For example, \citet{Bonnefoy2014a} find M9--L3 objects can match $\kappa$ And b's photometry; \citet{Hinkley2013} find a best fit with a L4 field dwarf and an intermediate gravity L1 dwarf (L1$\beta$).}.    Comparing the CHARIS spectra to both optically-anchored spectral templates and spectral libraries shows that $\kappa$ And b best resembles a young, low-gravity L0--L1 dwarf (L0--L1$\gamma$) like 2MASSJ0141-4633.  Its $H$-band spectral shape in particular shows strong evidence for a low surface gravity.

A number of factors may explain why our conclusions about $\kappa$ And b's spectrum show small differences with those presented in \citet{Hinkley2013}.   
Chiefly, the signal-to-noise ratio of $\kappa$ And b's spectrum is substantially higher (SNR$_{med., CHARIS}$ $\sim$ 20.5 vs. 5 for P1640), in large part owing to SCExAO's extremely high-fidelity AO correction resulting in a deep raw contrast.   This allowed us to extract a higher-fidelity spectrum and more clearly identify which spectral templates and empirical spectra match $\kappa$ And b.      Furthermore, calibrating the $\kappa$ And b spectrum from P1640 data is arguably more challenging since it relies on forward-modeling data reduced using SDI only \citep[see ][]{Pueyo2016}.   The slightly wider, redder bandpass (1.1--2.4 $\mu m$ vs. 0.9--1.8 $\mu m$) also probes more of $\kappa$ And b's spectral energy distribution, also aiding the identification of the companion's best-fit spectral properties.  Appendix B identifies an additional possible source of differences from the template spectrum used for spectrophotometric calibration.

While \citet{Carson2013} demonstrated that $\kappa$ And b is a bound companion, their short ($\sim$ 0.75 year) astrometric baseline precluded a detailed understanding of the companion's orbit, admitting a wide range of parameter space \citep{Blunt2017}.   Our astrometry establishes a 5-year baseline and decisively determined $\kappa$ And b's orbital direction (clockwise).   Orbital fits from two separate but complementary codes show that the companion's orbital plane is highly inclined relative to sky and possibly coplanar with the rotation axis of the star. Its eccentricity is likely substantial.   The semimajor axis of $\kappa$ And b suggests that the companion may orbit at a significantly wider separation than previously thought.    The companion's orbit -- including inclination and semimajor axis -- can be better clarified by including new astrometric measurements and determining solutions assuming observable-based priors \citep{KosmoONeil2018}.


\subsection{$\kappa$ And b in Context: Constraints/Limits on Temperature, Age, Gravity, Mass, and Formation}

While we reserve a detailed atmospheric modeling analysis of $\kappa$ And b for a future publication, we can use empirical comparisons
 to now quantitatively limit its temperature, revisit its age, and estimate its surface gravity and mass.   Combining these results with new information on $\kappa$ And b's orbit allows us to revisit a discussion of its plausible formation mechanisms.

\textbf{Temperature} -- A subset of the substellar objects whose spectra best fit $\kappa$ And b have a temperature derived from atmospheric modeling \citep{Bonnefoy2014b,Faherty2016}.      Conveniently, the best-fitting object -- 2MASSJ0141-4633 -- was analyzed in \citet{Bonnefoy2014a} using models incorporating cloud/atmospheric dust prescriptions that accurately reproduce young, early L dwarf spectrophotometry over 1--5 $\mu m$ \citep[][T. Currie et al. in prep.]{Daemgen2017}.   \citet{Bonnefoy2014a} derive $T_{\rm eff}$ = 1800$^{+200}_{-100}$ $K$.   While models utilized to constrain temperature in \citet{Faherty2016} were limiting cases that more poorly fit young, early L dwarfs, the derived temperature estimate for 2MASSJ0141-4633 using these models is consistent (1899 $K$ $\pm$ 123 $K$).   Temperatures for 2MASSJ0120-5200, 2MASSJ0241-5511 and 2MASSJ2322-6151B (all L1$\gamma$) are slightly lower, as expected, and consistent with the range of L0--L1$\gamma$ temperatures listed in \citet{Gonzales2018}.   Separately, temperatures for the closest-fitting field spectral type (L3) have a comparable range: \citep[1800--1900 $K$;][]{Stephens2009}.   Taken together, we estimate a temperature of 1700--2000 $K$ for $\kappa$ And b.

\textbf{Age} --  While a qualitative assessment of ``low gravity" generally means ``young", the mapping onto age may not be decisive.   Specifically, it is not clear yet how \textit{systematically} different substellar objects are in gravity class from $\sim$ 10 $Myr$ to 40 $Myr$ to 100 $Myr$, etc. and population studies may identify some overlap\footnote{For instance, while all good-fitting Tuc-Hor members are L0--L1$\gamma$, 
some L0--L1 objects in much older associations can also have a $\gamma$ designiation (e.g. AB Dor candidate member 2MASSJ2325-0259).   AB Dor includes likely members with both intermediate and low gravities at a given spectral type \citep{AllersLiu2013}.}.   Nevertheless, we can use properties of the best-fitting substellar objects coupled with system kinematics and interferometric measurements of the primary to determine whether multiple lines of evidence are consistent with the same \textit{likely} age of the $\kappa$ And system.   

 
 
  According to the Bayesian analysis tool for identifying moving group members, Banyan-$\Sigma$ \citep{Gagne2018a}, four of the seven objects in Table \ref{charisempcompare} are bona fide, decisive members of Tuc-Hor ($>$ 99.7\% membership probability), which has a Li-depletion age of 40$^{+5}_{-19}$ Myr \citep{Kraus2014}.   A fifth is a ``likely" member of Tuc-Hor (53\% probability) and sixth a possible member (25\% probability).   The other is a previously-identified candidate member of AB Dor (130--200 $Myr$) \citep{Bell2015}, where previous versions of Banyan (e.g. Banyan-II) estimated a far higher membership probability than does Banyan-$\Sigma$.    Tuc-Hor is comparable in age to the Columba  association \citep[t $\approx$ 30--40 $Myr$;][]{Zuckerman2011,Bell2015}, as both groups' pre-main sequences (luminosity vs. temperature) are nearly identical \citep{Bell2015}.   While $\kappa$ And b's proposed membership in Columba is highly suspect \citep{Hinkley2013}, using new GAIA-DR2 astrometry Banyan-$\Sigma$ still suggests it is a possible member (20\% probability)\footnote{Furthermore, the system's kinematics are identical to that of HR 8799 (50\% membership probability) within errors and its space position is similar.   Banyan-$\Sigma$ also does not consider ancillary information indicating that a particular system is young (e.g. spectral properties) -- $\kappa$ And is clearly not a Gyr-old system -- and new astrometry obtained with alternate kinematics codes may obtain different results \citep[e.g. ][]{Dupuy2018}. }.
 
Thus, regardless of whether $\kappa$ Andromedae actually is a member of Columba, properties of both the primary \textit{and} companion are consistent with what a system coeval with Columba should look like. 
Considering all lines of evidence together, we favor an age of 40$^{+34}_{-19}$ $Myr$, where the upper and lower bounds are equated with the age upper bound for the primary and the lower bound for most best-fitting comparison spectra, respectively\footnote{Taken at face value, this result appears to contradict that obtained by \citet{Hinkley2013}, who find that $\kappa$ And is likely at least 200 $Myr$ old.   However, as clearly stated in \citeauthor{Hinkley2013}, a much younger age is possible if the primary is a fast rotator viewed pole-on, which is exactly what was found in \citet{Jones2016}.  Thus, our two studies yield consistent answers on the system's age.}.
 

\textbf{Gravity} -- While there are few direct anchors for surface gravity for young substellar objects \citep[see][and T. Currie et al. 2018 in prep.]{Stassun2006,Stassun2007,Canty2013}, atmosphere/substellar evolution models can help identify plausible values for $\kappa$ And b.   Although a small subset of best-fit models that reproduced 2MASSJ0141-4633's spectrum in \citet{Bonnefoy2014b} had high surface gravities expected for field objects (log(g) $\sim$ 5--5.5), most had log(g) = 4.0 $\pm$ 0.5.    Using the \citet{Baraffe2003} evolutionary models, this object, siblings in Tuc-Hor, and slightly younger (20 Myr-old) ones are predicted to have surface gravities on the order of log(g) $\sim$ 4.1--4.2, while those of comparable temperature near our preferred upper age limit of $\sim$ 74 $Myr$ should have log(g) $\sim$ 4.5.   Surface gravities of log(g) $\sim$ 4--4.5 are therefore supported by a joint consideration of detailed atmosphere modeling of best-fitting spectra and predictions from evolutionary models covering $\kappa$ And's most plausible age range so far.
   
\textbf{Mass} -- Armed with a revised estimate for $\kappa$ And b's spectral type, photometry, and the system's distance, we calculate a bolometric luminosity of log(L/L$_{\rm \odot}$) = -3.81 $\pm$ 0.06 using the bolometric correction obtained by \citet{Todorov2010} for 2MASSJ0141-4633\footnote{Using the K-correction from \citet{Golimowski2004} for the best-fit field spectral type (L3) yields very similar results, consistent within errors (log(L/L$_{\rm \odot}$) $\sim$ -3.79).}.    Luminosities for the best-fitting L dwarfs in Tuc-Hor are comparable to $\kappa$ And b or slightly higher by 0.25 dex (-3.55 to - 3.8).   As their implied masses are 12--15 $M_{\rm J}$, if $\kappa$ And is coeval with Tuc-Hor then $\kappa$ And b is likely lower in mass.   Considering the full range of favored system ages, $\kappa$ And b's estimated mass is 13$^{+12}_{-2}$ $M_{\rm J}$ and companion-to-primary mass ratio is $q$ $\sim$ 0.005$^{+0.005}_{-0.001}$.

\textbf{Formation} -- Our results provide new information helpful for assessing how $\kappa$ And b relates to bona fide planets detected by both indirect techniques and direct imaging and low-mass brown dwarfs.   While the companion's mass is near or may even exceed the deuterium-burning limit, the utility and physical basis of this IAU criterion or any other hard mass upper limit for a ``planet" is unclear \citep{Luhman2008}\footnote{For example, the 2MASS J0441+2301 quadruple system \citep{Todorov2010,Bowler2015} includes two low-mass companions (M $\sim$ 10, 20 $M_{\rm J}$), suggesting that binary stars formed from molecular cloud fragmentation could still satisfy the IAU definition of a ``planet".}.   
Alternate criteria focusing on the demographics of imaged companions -- mass ratio and separation -- may more clearly distinguish planets from brown dwarf companions \citep{Kratter2010,Currie2011a}.  

While the plausible mass ratios of $\kappa$ And b are intermediate between that of HR 8799 cde ($q$ $\sim$ 4.5$\times$10$^{-3}$) and ROXs 42Bb ($q$ $\sim$ 9$\times$10$^{-3}$), its orbital separation is likely larger than any HR 8799 planets, more comparable to HIP 65426 b and ROXs 42Bb \citep[90--150 au;][]{Chauvin2017,Currie2014a}.   Similar to $\kappa$ And b, no additional companions have been found at smaller separations around HIP 65426 or ROXs 42B \citep{Chauvin2017,Bryan2016}.   Although core accretion struggles to form massive companions in situ beyond 50--100 au, disk instability may yet be a viable mechanism to account for $\kappa$ And b, HIP 65426 b, and ROXs 42Bb \citep[e.g.][]{Rafikov2005}.   At least some protoplanetary disks contain a significant amount of mass at 50--150 au-scale separations that could be (and perhaps have been) converted into massive companions via gravitational instability \citep[e.g.][]{AndrewsWilliams2007,Isella2016}, although direct imaging surveys show that superjovian-mass planets at these separations are rare \citep{Nielsen2013,Brandt2014,Galicher2016}.


\subsection{Future Studies of $\kappa$ And b}
Follow-up low-resolution CHARIS spectroscopy in individual passbands ($J$/$H$/$K$; $R$ $\sim$ 80) could better clarify $\kappa$ And b's atmospheric properties.     Gravity-sensitive indices $H$-cont and $H_{2}K$ approximated in this work could be more reliably determined; $J$ band potassium lines (K$_{I}$) could provide a third assessment of the companion's gravity \citep{AllersLiu2013}.    An improved census of substellar objects with ages at or just greater than that of Columba/Tuc-Hor (40--100 $Myr$) aided by the identification of new moving groups \citep[e.g.][]{Gagne2018b} could better establish a context for $\kappa$ And b and how its spectrum compares to the full range of very low, low, and intermediate gravity objects.
Ground-based broadband photometry can bracket CHARIS's coverage and also better probe evidence for clouds and small atmospheric dust, while more precisely constraining the companion's temperature \citep{Currie2011a,Currie2013,Daemgen2017}.  Thermal infrared observations with the \textit{James Webb Space Telescope} could reveal and help begin to quantify the abundance of $CO$, $CH_{4}$, and CO$_{2}$ \citep{BeichmanGreene2018}.

Higher-resolution ($R$ $\sim$ 3000) integral field spectroscopy of $\kappa$ And b achievable with Keck/OSIRIS and later on the \textit{Thirty Meter Telescope} with IRIS will provide a signficant advance in understanding $\kappa$ And b's gravity, clouds, chemistry, and perhaps formation \citep{Larkin2006, Larkin2016, Wright2014}.   OSIRIS and IRIS spectra can measure narrow gravity-sensitive lines of iron and sodium \citep{AllersLiu2013}.   Fitting these spectra with sophisticated forward models or analyzing them atmospheric retrievals should also yield estimates for $CO$, $H_{2}O$, $CH_{4}$, and perhaps NH$_{3}$ abundances from resolved molecular line emission \citep{Barman2015, Todorov2016}.    The carbon-to-oxygen ratio derived from these abundance estimates may provide insights into the formation environment of $\kappa$ And b and perhaps identifying with other directly-imaged planets \citep[e.g.][]{Barman2015}.

\acknowledgements The anonymous referee provided helpful comments, which improved the quality of this paper.    Eric Mamajek, Jonathan Gagne, Sasha Hinkley, Nienke van der Marel, Maxwell Service, and Jessica Lu also provided helpful suggestions and comments on earlier manuscript drafts.   We thank Adam Burrows for providing atmosphere models and Sasha Hinkley for sharing the P1640 spectrum for $\kappa$ And b.   We wish to emphasize the pivotal cultural role and reverence that the summit of Maunakea has always had within the indigenous Hawaiian community.  We are most fortunate to have the privilege to conduct scientific observations from this mountain and thus acknowledge the responsibility to use telescope time on Maunakea wisely.    This research has made use of the Keck Observatory Archive (KOA), which is operated by the W. M. Keck Observatory and the NASA Exoplanet Science Institute (NExScI), under contract with the National Aeronautics and Space Administration.   TC is supported by a NASA Senior Postdoctoral Fellowship.  SB is supported by a NSF Graduate Research Fellowship.  JC received support from the U.S. Fulbright program and from SC Space Grant.

{}

\appendix
\section{CHARIS Astrometric Calibration}
\begin{figure*}[ht]
\centering
\includegraphics[trim=25mm 0mm 25mm 0mm,clip,scale=0.275]{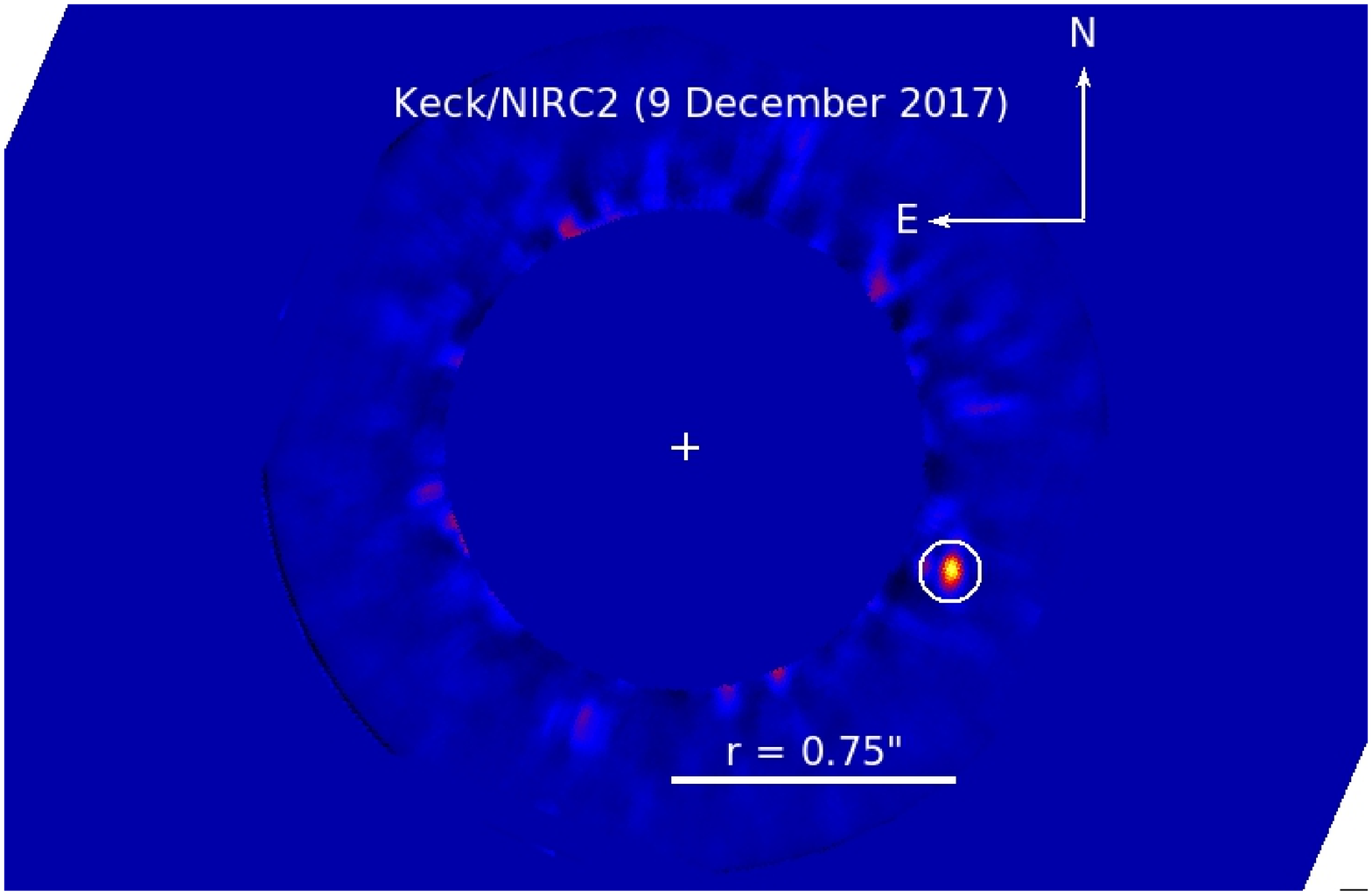}
\includegraphics[trim=25mm 0mm 25mm 0mm,clip,scale=0.275]{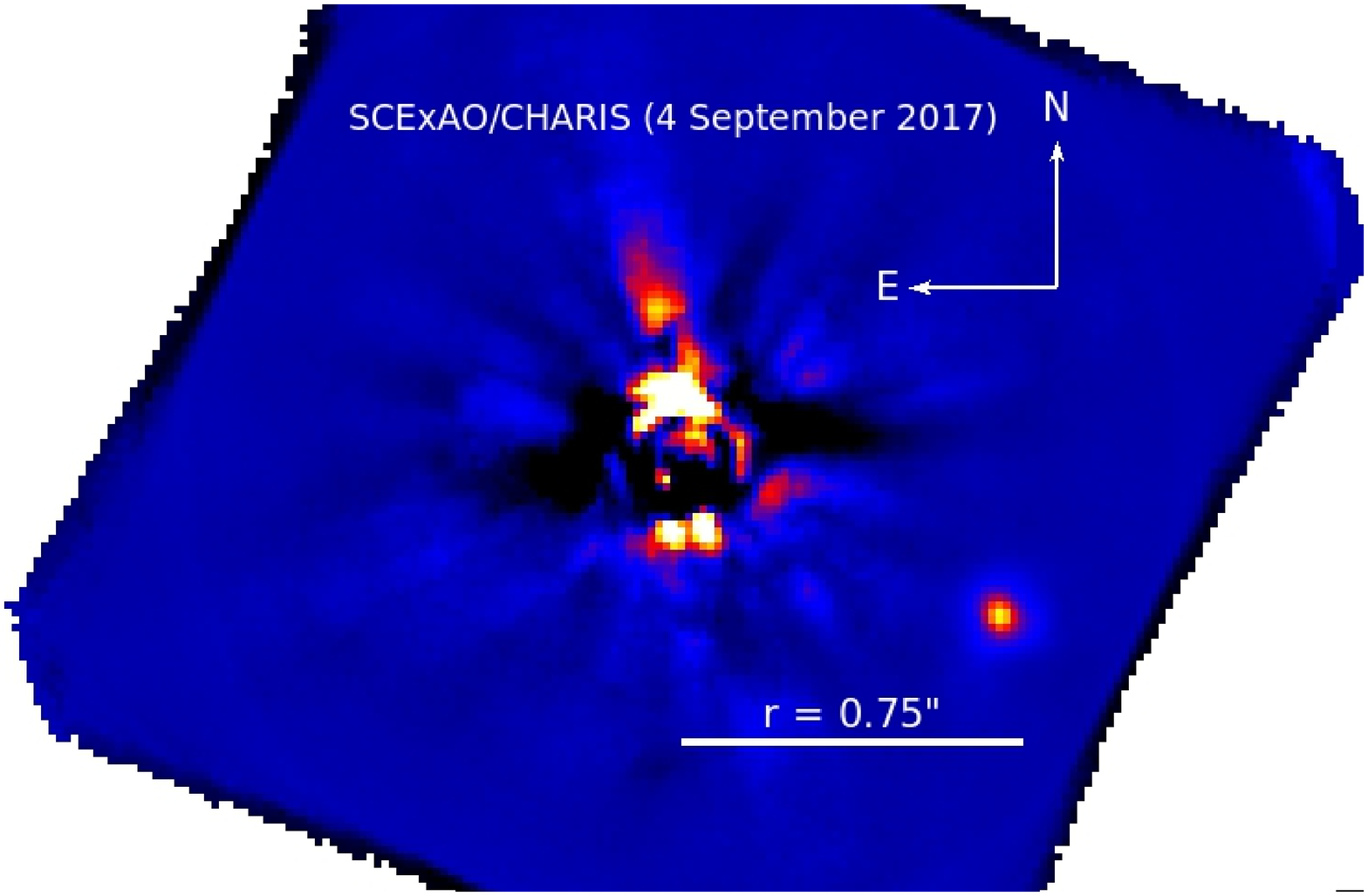}
\includegraphics[trim=25mm 0mm 25mm 0mm,clip,scale=0.275]{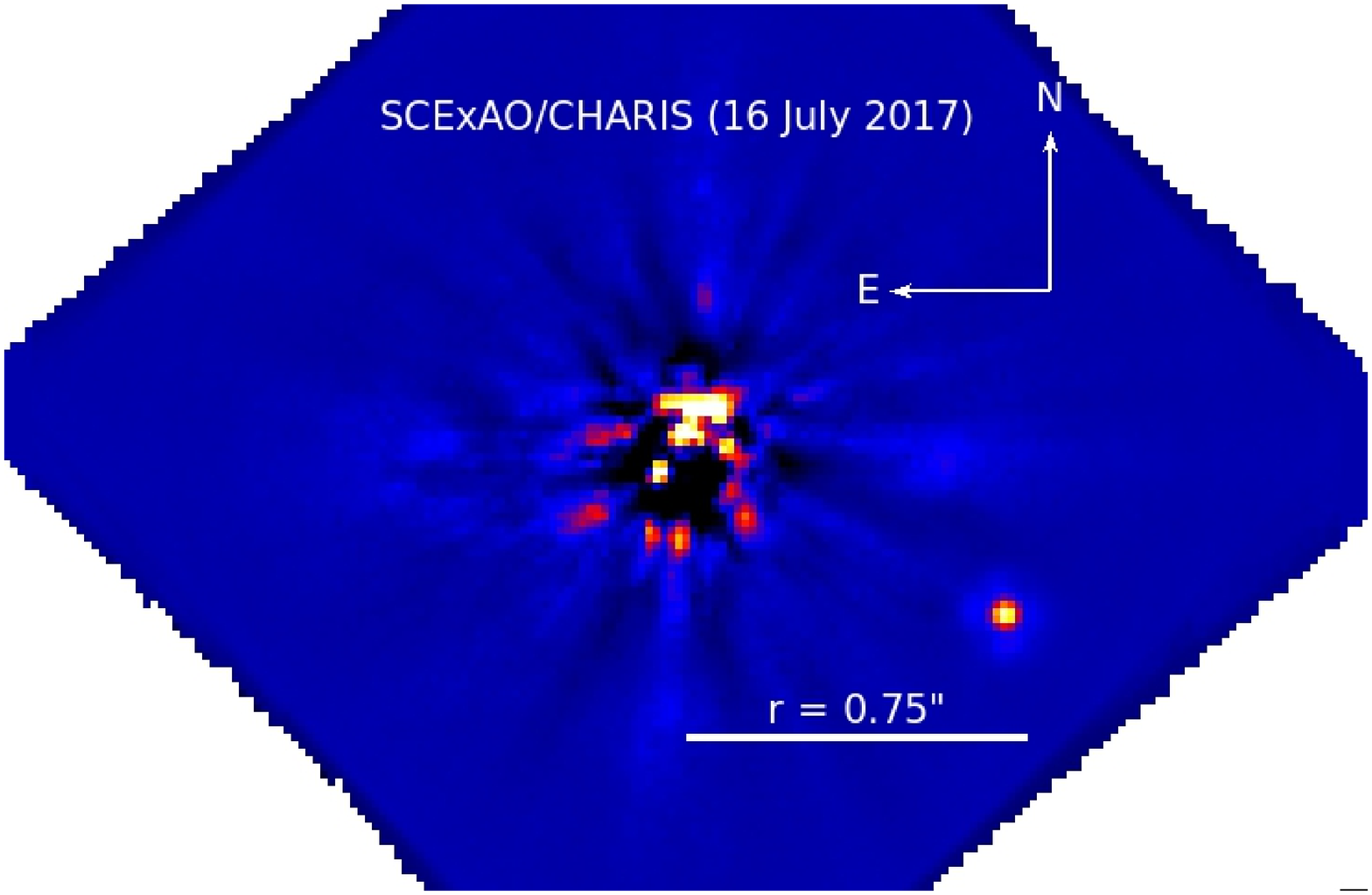}
\caption{New HD 1160 data used for a preliminary CHARIS astrometric calibration:  (left) Keck/NIRC2 from December 2017, (middle) SCExAO/CHARIS from 4 September 2017, and (right) SCExAO/CHARIS from 16 July 2017.   Previously published SCExAO/CHARIS data from 6 September used in our calibration are shown in \citet{Currie2018a}.}
\label{chariscalfig}
\end{figure*}
While precise astrometric calibration is ongoing, in this paper we present a preliminary calibration tied to Keck/NIRC2 based on July 2017, September 2017, and December 2017 observations the HD 1160 system.  HD 1160 has two wide (sub-)stellar companions \citep{Nielsen2012}, one of which (HD 1160 B) is near the edge of the CHARIS field of view at $\rho$ $\approx$ 0\farcs{}8.        At a projected separation of $r$ $\sim$ 80 au, the low-mass companion HD 1160 B should not experience significant orbital motion \citep{Nielsen2012, Garcia2017}.   Specifically, using the orbital fits from \citet{Blunt2017}, the separation and position angle for HD 1160 B are expected to change by $\Delta$$\rho$ $\sim$ -0.27 mas $\pm$ 0.36 mas, $\Delta$PA $\sim$ 0.026$^{o}$ $\pm$ 0.01$^{o}$  between September and December 2017 and $\Delta$$\rho$ $\sim$ -0.41 mas $\pm$ 0.54 mas , $\Delta$PA $\sim$ 0.040$^{o}$ $\pm$ 0.014$^{o}$ between July and December 2017.   At the separation of HD 1160 B a position angle change of 0.04$^{o}$ is no greater than $\sim$ 5\% of a NIRC2/CHARIS pixel: effectively HD 1160 B is stationary over this timeframe.

Keck/NIRC2 is precisely calibrated, with a north position angle uncertainty of 0.02$^{o}$ and post-distortion corrected astrometric uncertainty of 0.5 mas \citep{Service2016}.   Thus, we pinned the SCExAO/CHARIS astrometry for HD 1160 B to that for Keck/NIRC2 to calibrate CHARIS's pixel scale and north position angle offset.   This strategy follows that of the \textit{Gemini Planet Imager} campaign team in using contemporaneous GPI and Keck/NIRC2 imaging of HR 8799 to fine-tune GPI's astrometry \citep{Konopacky2014}.

Keck/NIRC2 $K$-band data for HD 1160 were obtained on UT 9 December 2017, immediately after $\kappa$ And, using the 0\farcs{}6 diameter partially transmissive coronagraphic spot.    Images consist of 11 coadded covering roughly 5 degrees in parallactic angle motion.    Basic NIRC2 data reduction procedures  -- flatfielding, dark subtraction, bad pixel mitigation, (post-rebuild) distortion correction, and image registration -- follows the pipeline from \citet{Currie2011a} previously used to process ground-based broadband data.   HD 1160 B was visible in the raw data; no PSF methods were applied.   However, the AO correction was modest and the star was blocked by coronagraph: we assumed a centroid uncertainty of 0.25 pixels in both directions.   In determining the error budget, we also considered the intrinsic SNR of the detection.

The $JHK$ data for HD 1160 from SCExAO/CHARIS data for HD 1160 were previously reported in \citet{Currie2018a}, taken on 6 September 2017 in two sequences, one with the Lyot coronagraph and another using the shaped-pupil coronagraph with good AO performance.   HD 1160 B is detected at a high significance in both data sets in all individual channels and data cubes, even without PSF subtraction techniques applied (SNR $\sim$ 100 in the wavelength-collapsed, sequence combined image).  To the astrometry extracted from these data, we add astrometry determined from 4 September 2017 (obtained under extremely poor conditions) and 16 July 2017 (obtained under excellent conditions).    Nominal astrometric errors consider the intrinsic SNR and a conservative estimate for the centroid (set to 0.25 pixels).

Table \ref{charisastromcal} shows our resulting astrometry for HD 1160 B; Figure \ref{chariscalfig} show images for NIRC2 data and previously unpublished SCExAO/CHARIS data.   For the nominal CHARIS astrometric calibration (0\farcs{}0164 pixel$^{-1}$ and no north position angle offset), the CHARIS astrometry displays no significant astrometric deviation between data sets but is systematically offset from the Keck/NIRC2 astrometry.   Taking uncertainty-weighted average astrometric offset, we obtain a revised pixel scale of 0\farcs{}0162 pixel$^{-1}$ $\pm$ 0\farcs{}0001 pixel$^{-1}$ and a north position angle offset of -2.20$^{o}$ $\pm$ 0.27$^{o}$ east of north (i.e. CHARIS data must be rotated an additional 2.2 degrees counterclockwise to achieve a north-up image).   

\input{charis_astrocal.tex}

\section{Absolute Spectrophotometric Calibration}
\begin{figure*}[ht]
\centering
\includegraphics[scale=0.5]{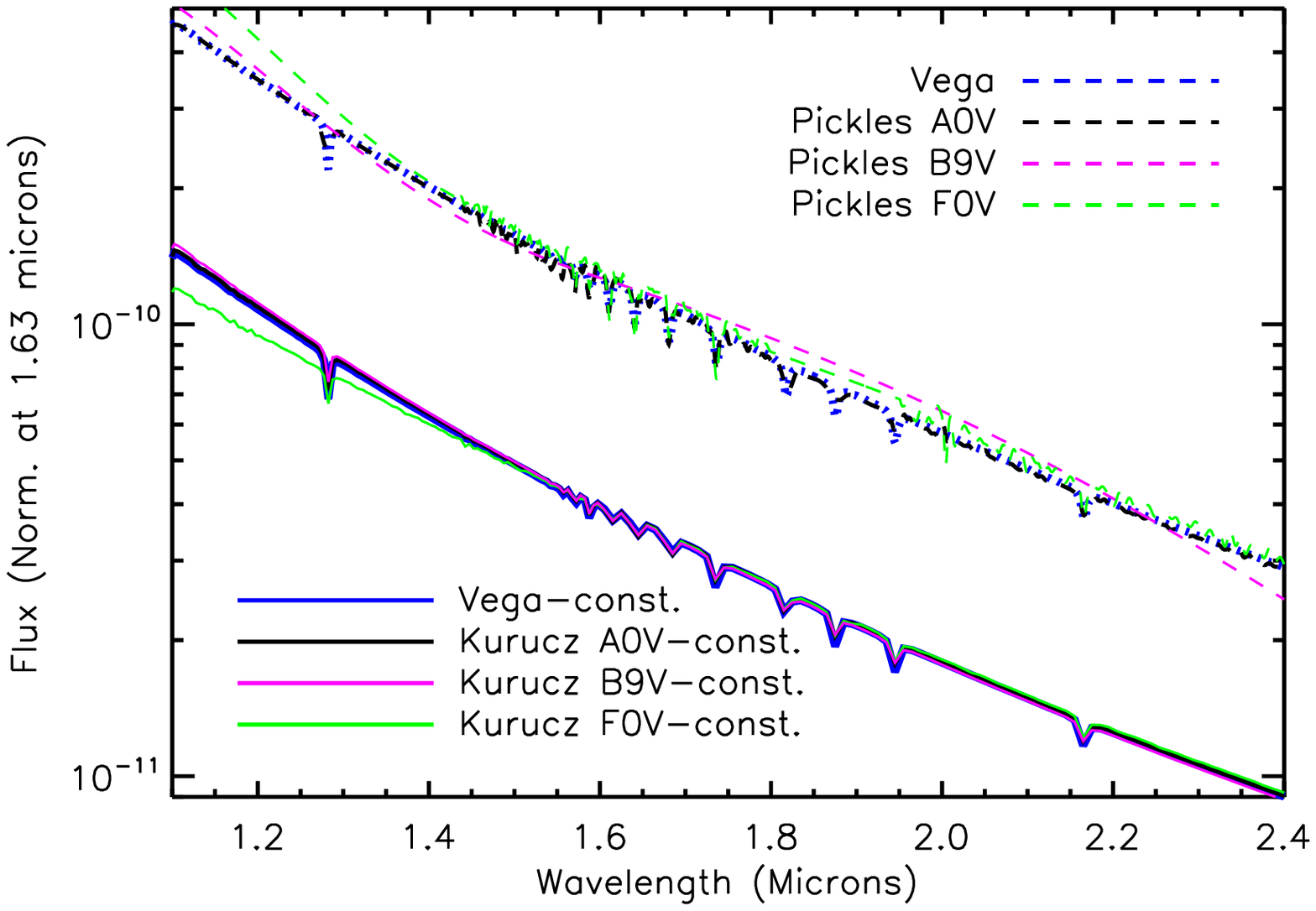}
\includegraphics[scale=0.5]{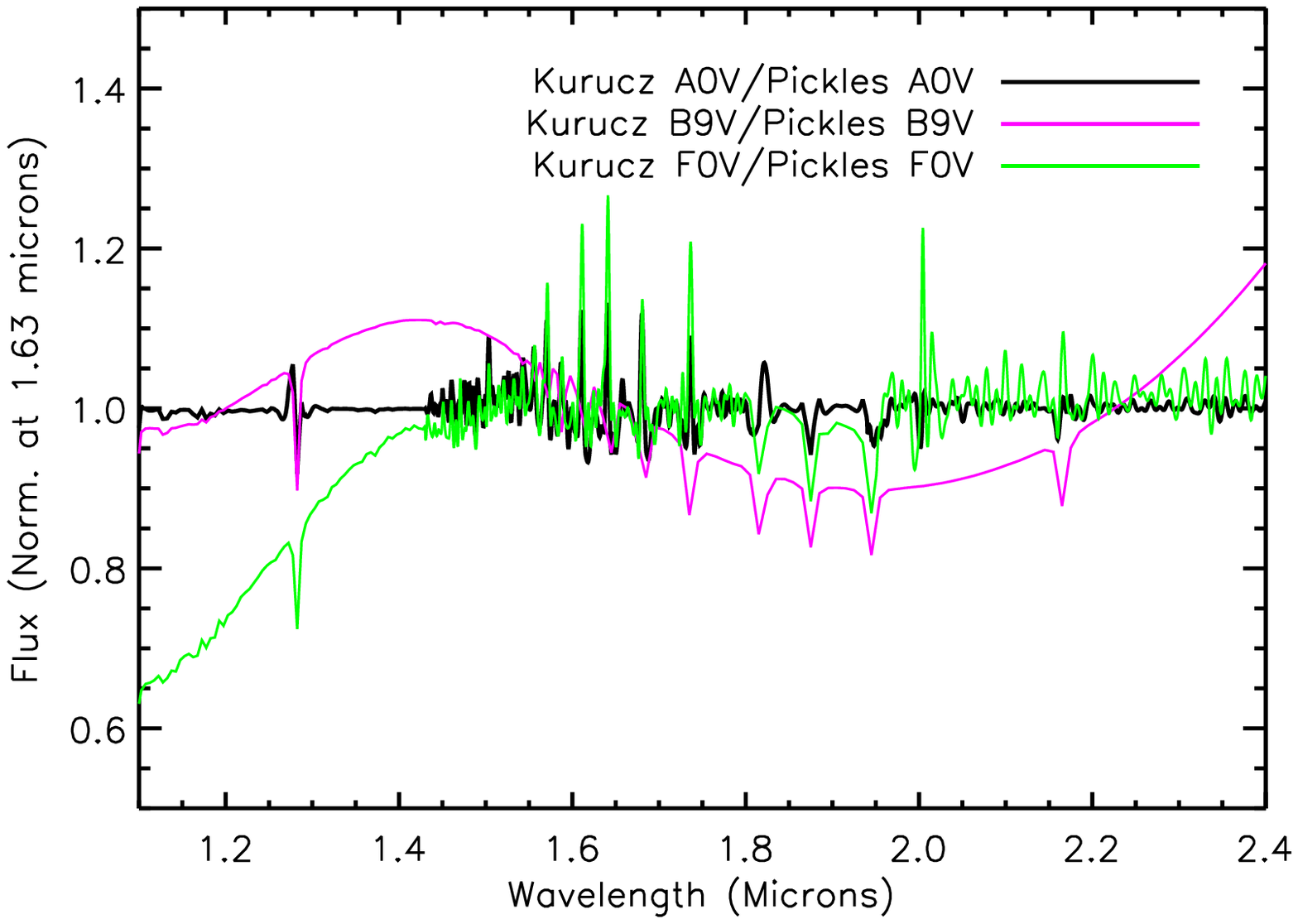}
\caption{(left) Near-infrared spectra from the Pickles library (dashed lines) and from the \citet{CastelliKurucz1993} atmosphere models (solid lines, offset by a constant).    The spectra are normalized for each source at 1.63 $\mu m$ and compared to Vega's spectrum.   Despite being of a very similar temperature, the B9V Pickles models exhibits significant offsets that are not predicted by the Kurucz models and are larger than a photosphere (F0V) whose temperature \citep[$\sim$ 7200 $K$;][]{Currie2010,Pecaut2012} is significantly cooler.   The Kurucz model spectra are nearly identical except at the shortest wavelengths for the coolest (F0V) model. (right) Ratio of the Kurucz to Pickles models for a given spectral type.   At B9V, the Pickles models induce errors in spectrophotometric calibration up to 20\%.}
\label{kuruczpicklescomp}
\end{figure*}

A key challenge with the new generation of coronagraphic extreme AO facilities is absolute spectrophotometric calibration.   Unocculted images of the star are often unavailable and satellite spots of a known attenuation are used to estimate a planet-to-star contrast in each spectral channel.
 \textit{Absolute} spectrophotometric calibration is necessary for accurate conclusions about any extracted planet/disk spectrum and requires an accurate model of the intrinsic spectrum of the unresolved target star (or a reference star) \citep[e.g.][]{Currie2017b}.
 
 The Pickles spectral library is the standard source for spectrophotometric calibration in the GPI Data Reduction Pipeline and has been used in direct imaging discovery and characterization papers \citep[e.g.][]{Macintosh2015}.   Importantly, it was used to calibrate P1640 spectra for $\kappa$ And b in \citet{Hinkley2013}.   However, we opted to use a robust, standard stellar atmosphere model \citep{CastelliKurucz1993} instead.  This is because \textit{we identified a potentially serious complication with multiple Pickles library entries at a level important for interpreting low-resolution planet/brown dwarf spectra.}. 
 
 Critically, \citet{Pickles1998} notes that near-infrared spectra is present for a few standard spectral types (e.g. A0V) but absent for the vast majority of their library, including the B9V spectral type.  For spectral types lacking near-IR spectra, \citeauthor{Pickles1998} uses ``a smooth energy distribution" extending beyond the reddest available wavelength (typically 1.04 $\mu m$) to 5 $\mu m$ such that the integrated broadband photometry in major near-IR passbands match published values.   However, this does not demonstrate that the spectral shape sampled at smaller $\Delta \lambda$ is consistent.
 
 Figure \ref{kuruczpicklescomp} compares B9V and A0V Pickles spectra and counterparts from the Kurucz atmosphere models.   The Pickles A0V, Kurucz A0V, and empirical Vega spectrum show strong agreement (left panel).   The ratio of the Kurucz A0V to B9V spectrum over the CHARIS passbands is nearly constant, as expected for two objects with similar temperatures and similar exponential terms in their Planck functions (e.g. at $\lambda$ = [1.25, 2.15] $\mu m$ this ratio is [1.27, 1.22]) and a lack of broad molecular absorption features.   Thus, we expect a very slowly changing or constant ratio of A0V/B9V over CHARIS passbands for the Pickles library spectra.    \textit{However}, as clearly shown in the right panel, the A0V/B9V flux ratio is unexpectedly variable over the CHARIS passbands, deviating by up to 20\% compared to the Kurucz atmosphere models and simple predictions based on pure blackbody emission.
 
 The practical consequence of using the Pickles B9V spectrum with extrapolated near-IR values instead of a stellar atmosphere model would be to suppress $\kappa$ And b's signal at 1.4 $\mu m$ and the red edge of K and increase it at $\sim$ 1.7--1.8 $\mu m$.  These wavelengths overlap with those sampled for gravity sensitive indices.   Thus, it is possible that some of our different results for the nature of $\kappa$ And b vs. \citet{Hinkley2013} are due to issues with the Pickles B9V spectrum which have only now been highlighted.   The choice of a proper stellar library may have important implications for interpreting substellar object spectra around other types of stars: for example, a $J$ spectrum extracted for a companion around an F0V star would deviate even more, perhaps leading to a misestimate of the companion's temperature.

\section{A Generalized, Robust Forward-Modeling/Spectral Throughput Calibration Using (A-)LOCI}
Powerful advanced least-squares PSF-subtraction algorithms like LOCI, KLIP, and derivatives can bias astrophysical signals, both reducing and changing the spatial distribution of the source intensity, thus affecting both spectophotometry and astrometry \citep[e.g.][]{Marois2010b,Pueyo2012}.   The earliest attempts at correcting for this annealing focused on injecting synthetic point sources at a given separation but different position angles and then processing real data with these sources added in successive iterations to estimate throughput \citep[e.g.][]{Lafreniere2007a}.   This approach yields a good estimate of the azimuthally-averaged point source throughput suitable for deriving contrast curves; however, it is computationally expensive \citep[e.g.][]{Brandt2013}.   Moreover, it is unsuitable for very precise spectrophotometry.   This is because algorithm throughput can vary at different angles at a given separation if the intensity of the stellar halo has a high dynamic range (e.g., if it is ``clumpy"), since high signal regions contribute more strongly to the residuals that the algorithm seeks to minimize \citep{Marois2010a}.   

\textit{Forward-modeling} provides a way to more accurately recover the intrinsic planet/disk brightness and astrometry/geometry, where the earliest methods focused on inserting negative copies of a planet PSF into the observing sequence with a brightness and position varied until it completely nulls the observed planet signal \citep{Marois2010a,Lagrange2010}.    With the planet signal entirely removed from the reference library used in these algorithms, PSF subtracted images containing the planet signal have 100\% throughput \citep{Currie2014b}.   While robust, this method is also computationally expensive for integral field spectrograph data instead of single band photometry (i.e. the runtime is $n_{channels}$ more lengthy) or if the intensity distribution of the signal is unknown (e.g. a disk of some morphology) \citep{Pueyo2016}.    To circumvent this problem, forward-modeling can be carried out in a more predictive fashion, where coefficients (for LOCI and derivatives) or Karhunen-Lo\`eve modes (for KLIP) used for PSF subtraction on science data are applied to empty images/data cubes containing only a synthetic planet or disk model \citep{Soummer2012,Esposito2014,Pueyo2014,Currie2015a}.    However, if the planet/disk signal is contained in the reference library used for PSF subtraction, as is usually the case for ground-based imaging, the signal itself can perturb the KL modes/coefficients \citep{Brandt2013,Pueyo2016}.   For KLIP, \citet{Pueyo2016} developed a robust, generalized solution solving this problem, modeling the planet/disk signal as inducing a small perturbation on the KL modes.  

Within the classic LOCI formalism\footnote{\citet{Pueyo2016} do describe how to apply their forward-modeling approach to LOCI but with a different linear algebra formalism than utilized in \citet{Lafreniere2007a} and nearly all subsequent LOCI-based works, including \citet{Brandt2013}.},  \citet{Brandt2013} thus far has developed the advanced approach most similar to that done by \citet{Pueyo2016}, efficiently modeling the planet as inducing a small perturbation on LOCI coefficients, $\boldsymbol  \beta$.   Their focus was an efficient and rapid computation of contrast curves for broadband data, not precise spectrophotometry/astrometry.   In order to precompute the effective planet PSF inducing this perturbation, they therefore approximated it as a gaussian profile flanked by two sets of gaussian profiles at fixed position angle offsets and worked in the limiting case that the planet/disk intensity in the subtraction zone $I_{i,s}^{\prime}$ (not optimization zone) is far lower than the speckle intensity $I_{i,s}$.   Absent this robust approach, several authors have introduced must incorporate modifications like local masking, a very large optimization zone, or an aggressive singular value cutoff (large $svd_{lim}$) to substantially reduce the influence of perturbations LOCI-like algorithms \citep[e.g.][]{Marois2010b,Currie2015a}.   However, we found that these modifications alone failed to yield a high-signal-to-noise spectrum for $\kappa$ And b whose throughput is precisely known.
 
 Here, we develop a generalized forward-modeling solution in the (A-)LOCI formalism complementary to the KLIP forward-modeling approach in \citet{Pueyo2016}, adopting the formalism of and leveraging upon advances made by \citet{Brandt2013}.   In the standard case for LOCI-like algorithms, the set of coefficients $\alpha_{ij}$ applied to reference images $I_{j}$ determined from an optimization region $o$ minimize the subtraction residuals ${\cal R}_{i,o}^{2}$ over pixels $k$ for science image $I_{i}$:
 \begin{equation}
{\cal R}_{i,o}^{2} = \sum_k (I_{ik,o} - \sum_j \alpha_{ij} I_{jk,o}~)^{2}.
\label{eq:loci_opt}
\end{equation}

Following \citet{Brandt2013}, we can perturb this equation by adding a planet PSF of signal $I_{}^{\prime}$ to each reference image $j$ over pixels $k$, inducing a small perturbation in coefficients $\beta_{ij}$, thus the subtraction residuals are now:
 \begin{equation}
{\cal R}_{i,o}^{2} = \sum_k [I_{ik,o} + I_{ik,o}^{\prime} - \sum_j (\alpha_{ij} I_{j,o} + \alpha_{ij} I_{jk,o}^{\prime} + \beta_{ij}I_{jk,o}+\beta_{ij}I_{jk,o}^{\prime})~]^{2}, 
\label{eq:loci_optpert}
\end{equation}
Linearizing now around $\beta_{ij}$ instead of $\alpha_{ij}$, we find:
\begin{equation}
\frac{\partial {\cal R}_{i,o}^{2}}{\partial \beta_{il}} = 0 = \sum_k [(I_{lk,o} + I_{lk,o}^{\prime})(I_{ik,o} + I_{ik,o}^{\prime} - \sum_j (\alpha_{ij} I_{jk,o} + \alpha_{ij} I_{jk,o}^{\prime} + \beta_{ij}I_{jk,o}+\beta_{ij}I_{jk,o}^{\prime}))]
\end{equation} 

Over the optimization zone $o$ (not the subtraction zone) and summed over pixels $k$, we assume that the speckle halo intensity is much larger than a planet ($I_{i,o}$ $>>$ $I_{i,o}^{\prime}$) and that the speckle halo subtracted by the reference library weighted by the nominal (A-)LOCI coefficients is zero ($I_{i,o}$ $-$  $\sum\limits_j (\alpha_{ij,o} I_{j,o}$) $\sim$ 0), reducing to a system of linear equations,
\begin{equation}
\sum_k I_{lk,o} (I_{ik,o}^{\prime} - \sum_j \alpha_{ij} I_{jk,o}^{\prime}) = \sum_j \beta_{ij} \sum_k I_{lk,o} I_{jk,o},
\end{equation}
which can be solved by matrix inversion as done in Equation \ref{eq:invert},
\begin{equation}
 \boldsymbol{\beta}=(\mathbf{U}\mathbf{\Sigma_{> \mathit{svd_{lim}}}}\mathbf{V})^{-1}\cdot\mathbf{b^{\prime}},
  \label{eq:invertpert}
  \end{equation}

where \textbf{U}\textbf{$\Sigma$}\textbf{V} ($\boldsymbol{\rm A}$) is the same covariance matrix whose array elements are described in Equation \ref{eq:axbarrel} and  $\boldsymbol{\rm b^{\prime}}$ is the column matrix describing how the effective (partially annealed) planet PSF induces a set of perturbation with coefficients $\boldsymbol{\rm \beta}$:
\begin{equation}
{\rm b}_l^{\prime} =\sum_k I_{lk,o} (I_{ik,o}^{\prime} - \sum_j \alpha_{ij,o} I_{jk,o}^{\prime}).
\label{eq:axbpert}
\end{equation}

The residual emission of planet in frame $i$ within subtraction zone $s$ is then
\begin{equation}
{\cal R}_{i,s}^{\prime}= \sum_k [I_{ik,s}^{\prime} -  \sum_j (\alpha_{ij,o} I_{jk,s}^{\prime} + \beta_{ij,o}I_{jk,s}+\beta_{ij,o}I_{jk,s}^{\prime})].
\end{equation}

\begin{figure*}
\centering
\includegraphics[trim=10mm 0mm 0mm 0mm,clip,scale=0.225]{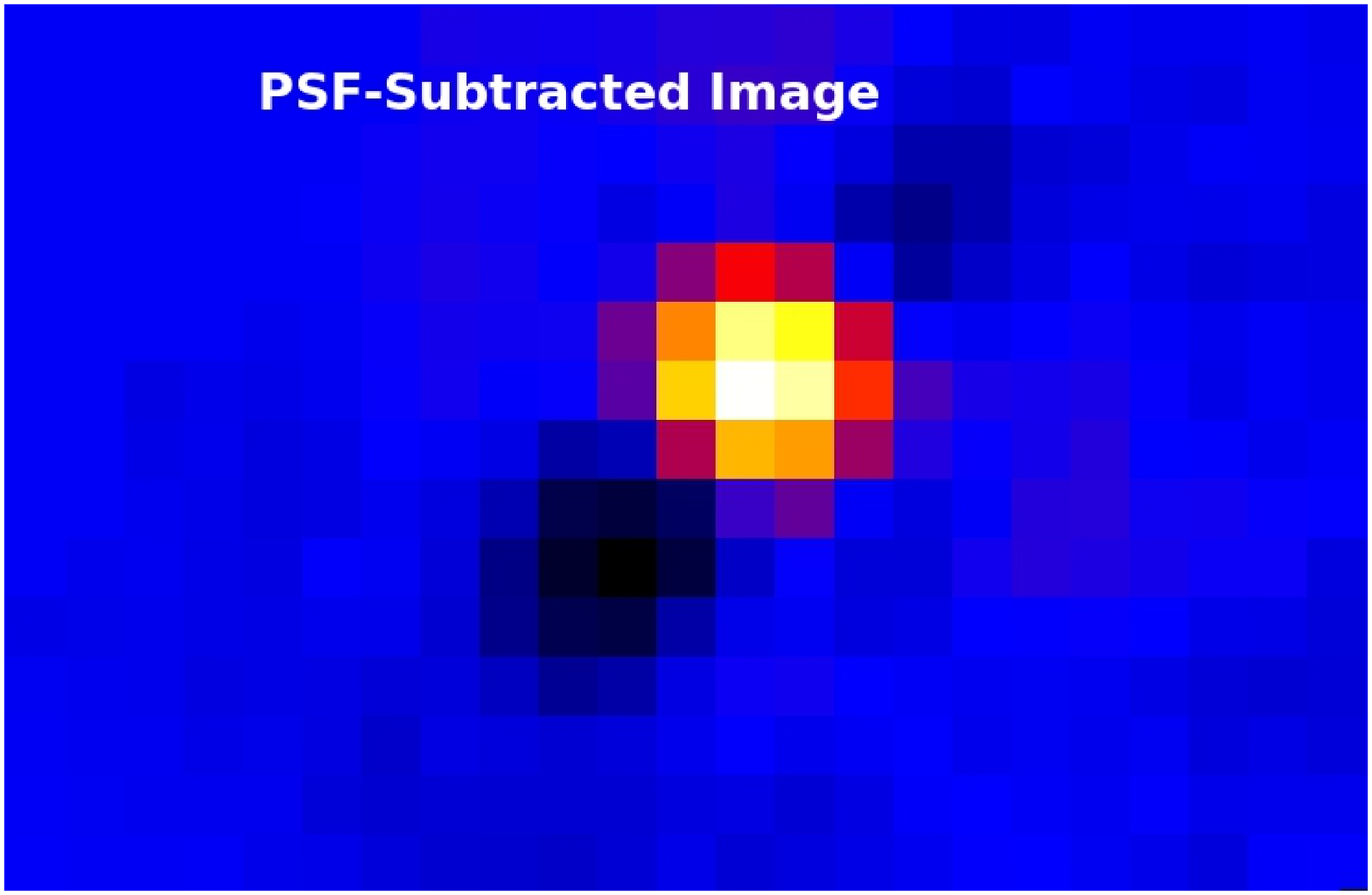}
\includegraphics[trim=10mm 0mm 0mm 0mm,clip,scale=0.225]{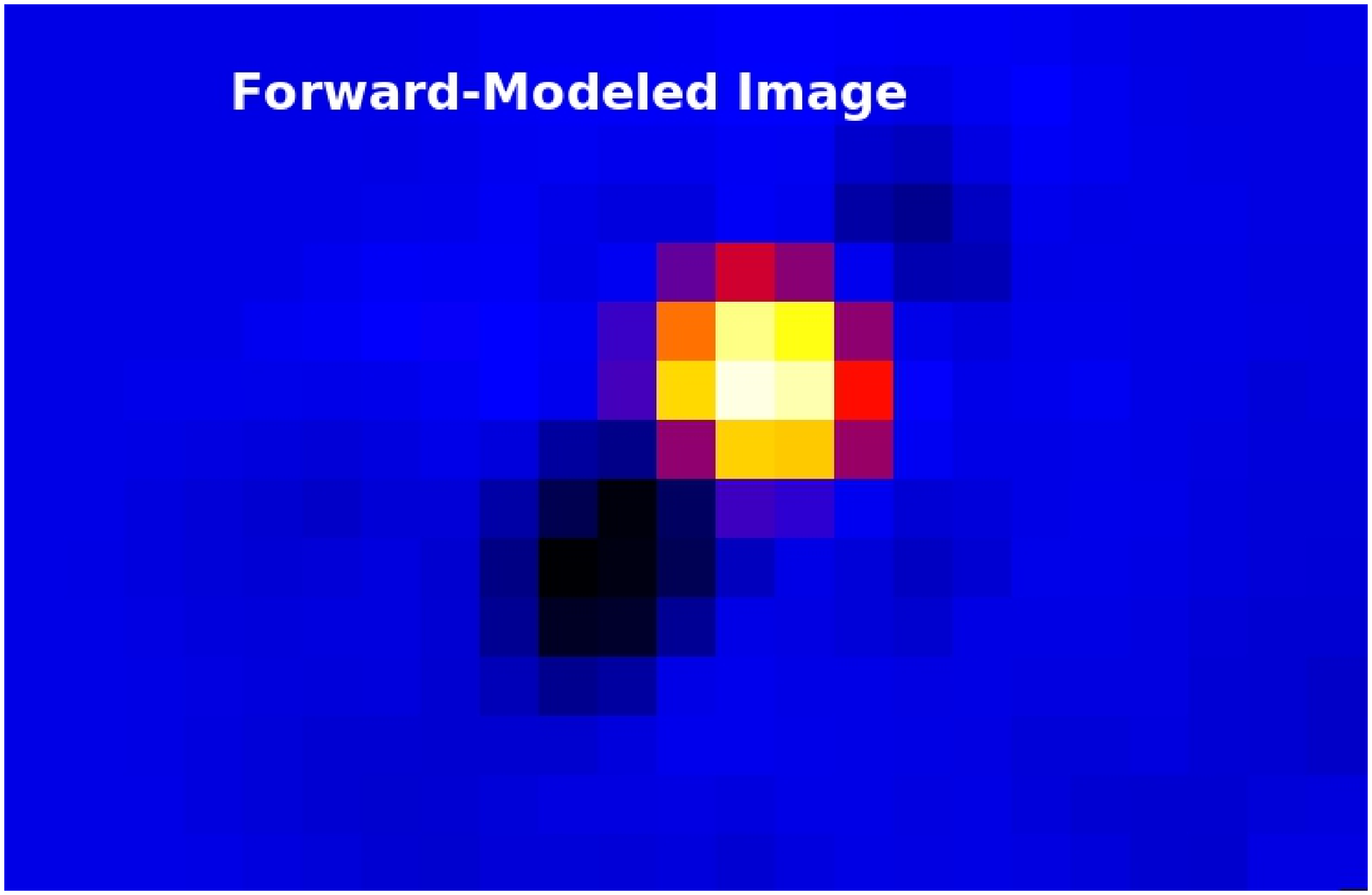}
\includegraphics[trim=10mm 0mm 0mm 0mm,clip,scale=0.225]{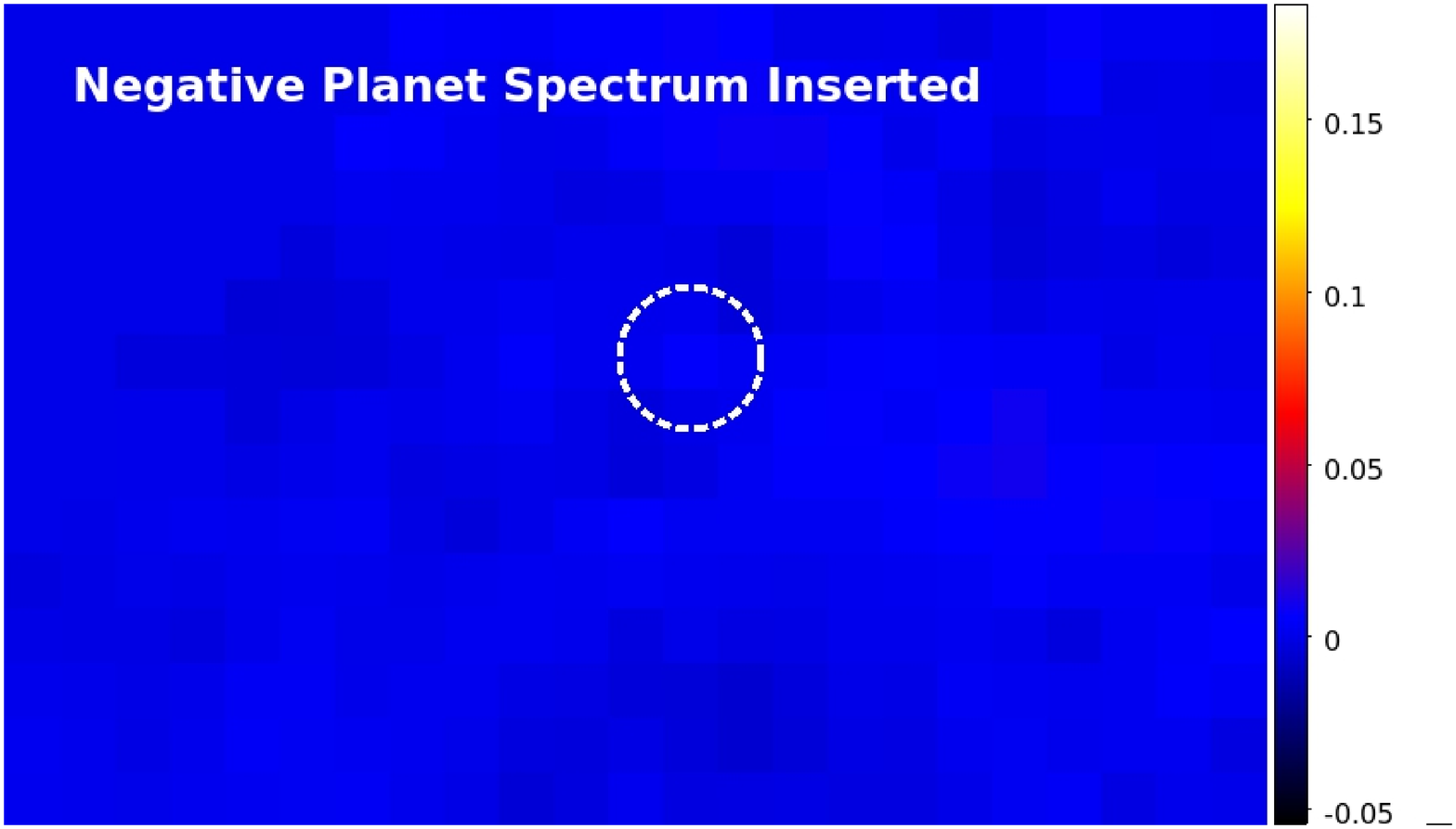}
\caption{Demonstration of our spectrophotometric throughput correction using a combination of forward-modeling and iterative nulling.   (left) The wavelength-collapsed PSF-subtracted image from Figure 1a showing clearly the signal peak from $\kappa$ And b and the negative subtraction footprints.   (middle) The forward-modeled image at roughly the location of $\kappa$ And b predicted from perturbed A-LOCI coefficients and used to estimate the wavelength-dependent throughput correction for the companion's spectrum.   (right) The PSF-subtracted image after inserting a negative copy of $\kappa$ And b's spectrum at the companion's location into each data cube in our observing sequence.   In all panels, the color stretch is the same and is in units of mJy (vertical bar).  The negative spectrum - derived from forward-modeling - almost perfectly nulls the planet signal within one resolution element (dashed circle).}
\label{alocifwdmoddemo}
\end{figure*}

Throughout, we use the satellite spots to produce a model planet PSF for each channel.   The empty data cubes containing the annealed planet PSF produced from the forward-model are then derotated and combined.   Their photometry and astrometry are compared to input values to derive throughput for each spectral channel and astrometric offsets.     These offsets were then applied to our spectral extraction subroutine to derive a flux-calibrated CHARIS spectrum.  To confirm the validity of our forward-model, we inserted a negative copy of the extracted, corrected CHARIS spectrum into the real data at its predicted position in each data cube $\boldsymbol{x}$ -- $I_{i,\lambda, neg.} = I_{i,\lambda}-I_{i,\lambda}^{\prime\prime}(\boldsymbol{x})$ --  and slightly varied the position/brightness of the spectrum to verify that the forward-modeled solution is the correct one.   

Figure \ref{alocifwdmoddemo} illustrates our forward-modeling method, demonstrating that the predicted annealed planet PSF matches the real one and the extracted CHARIS spectrum nulls the planet signal in the sequence-combined data cube.   For our $\kappa$ And b data set using algorithm parameters adopted in \S 2. the spectrum throughput ranges between 77\% and 92\%; the astrometric offset is $\sim$ 0.2 pixels.   In classic LOCI, very small optimization zones like those we adopt combined with poor field rotation can result in low throughput \citep{Lafreniere2007a}.   Our throughput is high and perturbations of the (A-)LOCI coefficients are low in large part because of local masking \citep[see also][]{Currie2013}.   This is because pixels corresponding to the subtraction zone $s$ (ostensibly containing most of the planet signal) are removed from the optimization zone $o$.   Since the perturbed coefficients $\boldsymbol \beta$ are determined from a system of linear equations considering the optimization zone only, their values are much smaller when local masking is used.   Truncating the covariance matrix $\textbf{A}$ also reduces the algorithm aggressiveness and potentially the planet's perturbations of (A-)LOCI coefficients.

This approach introduces some key modifications to that first proposed by \citet{Brandt2013} and utilized in the ACORNS pipeline.   First, we explicitly calculate the effective planet PSF in each frame $i$ -- $I_{ik,o}^{\prime} - \sum_j \alpha_{ij,o}$ -- since the angular displacements of PSFs in frames $j$ and the coefficients $\alpha_{ij,o}$ are unique for each data set.    Second, while we assumed that $I_{i,o}$ $>>$ $I_{i,o}^{\prime}$ to determine $\boldsymbol  \beta$, the planet signal over the (typically much smaller) subtraction zone may not always be negligible in each spectral channel.   Third, ACORNS (developed for low-Strehl data sets) modeled the planet PSF with a gaussian intensity distribution, whereas we use an empirical model unique for each data set.   Fourth, it explicitly incorporates the distinction between optimization and subtraction zones and incorporates local masking and covariance matrix truncation, although ACORNS can easily be modified to do this as well.

\section{Extracted SCExAO/CHARIS Spectrum for $\kappa$ And \lowercase{b}}
We provide our extracted SCExAO/CHARIS spectrum for $\kappa$ And b in Table \ref{chariskapandspecapp}.
\input{extracted_spectrum.tex}

\end{document}

%% file: cruz_spectralstandard.tex
\begin{deluxetable*}{lllllll}
\tablecaption{Fits to Cruz et al. (2018) Spectral Standards}
\tablehead{{Spectral Type} & {Gravity Class} & {H$_{\rm cont, CHARIS}$} & {H$_{2}$K, CHARIS} & {$\chi_{\nu}^{2}$ (total)} & {$\chi_{\nu}^{2}$ ($H$+$K$)}}
\tiny
\startdata
L0 & field & 0.935 & 1.050 & 3.76 & 3.72\\
L0 & $\beta$ & 0.945& 1.032 & 1.68 & 2.41 \\
L0 & $\gamma$ & 0.971 & 1.020 & \textbf{1.26} & \textbf{1.40}\\
L1 & field & 0.912 & 1.056 & 2.90 & 3.55\\
L1 & $\beta$ & 0.926 & 1.056 & 1.80& \textbf{1.71}\\
L1 & $\gamma$ & 0.949 & 1.037 & 2.84 & \textbf{1.41}\\
L2 & field & 0.896 & 1.076 & 2.03 & 2.44 \\
L2 & $\gamma$ & 0.960 & 1.009 & 5.10 &3.32 \\
L3 & field & 0.890 & 1.075 & \textbf{1.51} & \textbf{1.78} \\
L3 & $\gamma$ & 0.947 & 1.031 & 3.50&1.91 \\
L4 & field & 0.867 & 1.075 & 2.28 & 1.86 \\
L4 & $\gamma$ & 0.940 & 1.037 & 15.32 &10.96 \\
L6 & field & 0.847 & 1.110 & 3.36 & 2.94 \\
L7 & field & 0.855& 1.109 & 5.92 & 3.19 \\
L8 & field & 0.794& 1.172 & 5.00 &6.63\\
 \enddata
 \tablecomments{The $\chi_{\nu}^{2}$ values are calculated assuming 15 degrees of freedom for fitting of the JHK peaks and 10 for just $H$ and $K$.   Entries in bold identify those that fit the data to within the 95\% confidence limit.}
\label{charisspecstandard}
\end{deluxetable*}

%% file: empcompare_bestfit.tex
\begin{deluxetable*}{llllllllllll}
\tablecaption{Properties of the Best-Fitting Substellar Objects}
\tablewidth{0pt}
\tablehead{\colhead{Name} & \colhead{$\chi^{2}_{\nu}$} & \colhead{$\chi^{2}_{\nu}$}&\colhead{SpT} & \colhead{H$_{\rm cont.}$} & \colhead{H$_{2}$K} & \colhead{Assoc.} & \colhead{Age} & \colhead{log(L/
L$_{\odot}$)}&\colhead{$T_{\rm eff}$ (K)} & \colhead{log(g)} & \colhead{Mass (M$_{\rm J}$)} \\
{} & {(Total)} & {(H+K)} & {}& {Index} & {Index} & {} &{ (Myr)}& {(Approx.)} &{}&{(Approx.)}}
\tiny
\centering
\startdata
2MASSJ0141-4633 & \textbf{1.43} & \textbf{1.81} & L0-L1$\gamma$ & 0.962 & 1.027 & Tuc-Hor & 40$^{+5}_{-19}$ & -3.58 & 1899 $\pm$ 123 & 4.1--4.2 & 13--15  \\
   &  & & & & & & & & 1800$^{+200}_{-100}$\\
2MASSJ0120-5200 & 2.25 & 2.24 & L1$\gamma$ & 1.032 & 1.049 & Tuc-Hor & 40$^{+5}_{-19}$ & -3.65 & 1685 $\pm$ 145 & 4.1--4.2 & 12.5--14  \\
 2MASSJ0241-5511 & 1.83 & 2.59 & L1$\gamma$ & 1.015 & 1.034 & Tuc-Hor & 40$^{+5}_{-19}$& -3.67 & 1731 $\pm$ 151 & 4.1--4.2 & 12.5--14\\
2MASSJ0440-5126 & {1.79} &2.64& L0$\gamma$  & 1.003 & 1.006 & Tuc-Hor? (53) & 40$^{+5}_{-19}$? &-3.63? & 1600--2000 & 4.1--4.2? & 13--15? \\ 
2MASSJ2033-5635 & 1.71 &2.44  & L0$\gamma$ & 0.945& 1.034 & Tuc-Hor??$^{a}$ & ?? & ??&1600--2000 &?? &??\\
2MASSJ2325-0259 & \textbf{1.45} & 2.03 & L1$\gamma$ & 1.040 & 1.067 & AB Dor? (65) & 130--200?& -3.80?$^{b}$ &1700--1900 &4.7--4.9?&30--40?$^{b}$\\
2MASSJ2322-6151B & 1.94  & 2.26 & L1$\gamma$ & 1.015 & 1.083 & Tuc-Hor & 40$^{+5}_{-19}$ & -3.68 & 1793 $\pm$ 50 &4.1--4.2 &12.5-14\\
 \enddata
 \tablecomments{ 
 Spectra for all objects match $\kappa$ And b's at 99.7\% confidence for the $JHK$ restricted fit, the $HK$ restricted fit, and the $JHK$ unrestricted fit.
 Secure moving group members are defined from Banyan-$\Sigma$ as those with $>$ 95\% probability in a given group.   Those with $>$ 50\% are noted with "?": the Banyan-$\Sigma$ probability is listed in parentheses.       Temperatures are listed from \citet{Faherty2016} (first entry) or \citet{Bonnefoy2014a} (second entry) where available; otherwise, they are estimated from the range in temperatures from \citet{Gonzales2018}.     If given, luminosities, surface gravities, and masses are calculated assuming the nominal object distance, the K-band bolometric correction from \citet{Todorov2010}, and the \citet{Baraffe2003} luminosity evolution models.    a) Previously identified as a Tuc-Hor member, Banyan-$\Sigma$ favors a field object ($\sim$ 75\% vs. 25\%).   No parallax is given.   Thus its membership and properties depending on distance are noted with a "??". b) Mass and luminosity estimated using the ``optimal" kinematic distance for moving group membership.   
 }
 \label{charisempcompare}
\end{deluxetable*}

%% file: orbitfit.tex
\begin{deluxetable*}{llllll}
\tablecaption{Orbit Fitting for $\kappa$ And b}
\tablehead{ {} & {} & {\textbf{OFTI}} & {}& {\textbf{ExoSOFT}} \\ {Orbital Element} & {Unit} & {Median [68\% C.I.]} & {[95\% C.I.]} & {Median [68\% C.I.]} & { [95\% C.I.]}}
\tiny
\startdata
a & au & 76.5 [56.7, 128.2] & [47.2, 286.6]& 99.0 [53.7, 126.6] & [45.1, 216.1]\\
P & yr & 399.9 [254.9, 868.1] &[193.3, 2899.5]& 588.8 [214.1, 825.9] &[169.0, 1868.8]\\
e & & 0.80 [0.67, 0.87] &[0.54, 0.93]& 0.69 [0.59,0.83]& [0.47, 0.90]\\
i & $^{o}$ & 136.2 [119.6, 157.4]& [111.1, 171.5] & 121.2 [109.2,129.2] &[105.5, 158.7]\\
$\omega$ & $^{o}$  & 126.5 [49.1, 161.0]& [3.6, 176.7]& 129.5 [95.7, 157.2] &[71.1, 195.5]\\
$\Omega$ & $^{o}$ & 75.9 [54.1, 100.5]& [15.4, 162.1]& 75.7 [64.1, 87.0] &[31.1, 113.4]\\
$T_{0}$ & yr & 2042.7 [2039.1, 2051.5] &[2037.3, 2062.7]& 2047.62 [2038.38,2053.82] &[2036.14, 2069.47]\\
 \enddata
 \label{orbitfits}
 \tablecomments{Orbits are fit to the four new NIRC2 and CHARIS astrometric points plus two HiCIAO epochs listed in \citet{Carson2013}.}
\end{deluxetable*}

%% file: kapand_prop.tex
\begin{deluxetable*}{llll}

\tablecaption{Properties of the $\kappa$ And system}
\tiny
\tablehead{{Parameter} & {$\kappa$ And A} & {$\kappa$ And b} & {Reference}}
\centering
\startdata
\hline
Object Properties\\
\hline
d	(pc)										&		$50.0\pm0.1$		&		\dots		&	1		\\					
Age (Myr)										&		$47^{+27}_{-40}$	&	$\approx$ $40^{+34}_{-19}$? &		2, 6	\\	
Mass										& 		2.8 $M_{\odot}$ 	&      $\approx$ 13$^{+12}_{-2}$ $M_{\rm J}$? & 2, 6\\	
log(L/L$_{\odot}$)								& 		$1.80^{+1.7}_{-0.04}$ &    -3.81$\pm$ 0.05 & 2, 6\\
Spectral type								&		B9IV/B9V		&	L0-L1$\gamma$		&		2, 5, 6			\\
$\mathrm{T_{eff}}$ (K)	&   	$11 327^{+421}_{-44}$		&	 $1700-2000$   	&	 2, 6			\\	
log g		(dex)			    &		$4.174^{+0.019}_{-0.012}$     &     $\approx$ $4.0-4.5$?                      &   2, 4  \\		
\\
Photometry\\
\hline				
J	(mag)									&		$4.26\pm0.04$		&	$15.84\pm0.09$		&	2			\\
H	(mag)											&	$4.31\pm0.05$	&	$15.01\pm0.07$			&	2		\\
$\mathrm{K_{s}}$ (mag)						& 	$4.32\pm0.05$	&		$14.37\pm0.07$	&	2			\\
L'		(mag)										&	$4.32\pm0.05$ & 	$13.12\pm0.1$		&	3, 4			\\
NB\_4.05	(mag)								&		$4.32\pm0.05$		&	$13.0\pm0.2$		&	 4		\\
M'		(mag)										&	$4.30\pm0.06$	&	$13.3\pm0.3$		&	 4			\\
\hline
                                                \\
Astrometry\\
\hline
UT Date & Data Source & [E, N]\arcsec{}\\
\hline
2012 01 01.     & AO188/HiCIAO & [0.884 $\pm$ 0.010, 0.603 $\pm$ 0.011] & 3\\
2012 07 08      & AO188/HiCIAO & [0.877 $\pm$ 0.007, 0.592 $\pm$ 0.007] & 3\\
2012 11 03      & Keck/NIRC2 & [0.846 $\pm$ 0.010, 0.584 $\pm$ 0.010] & 2\\
2013 08 18      & Keck/NIRC2 & [0.829 $\pm$ 0.010, 0.585 $\pm$ 0.010] & 2\\
2017 09 05      & SCExAO/CHARIS & [0.710 $\pm$ 0.012, 0.576 $\pm$ 0.012] & 2\\
2017 12 09      & Keck/NIRC2 & [0.699 $\pm$ 0.010, 0.581 $\pm$ 0.010] & 2\\
 \enddata
 \tablecomments{References -- 1) Gaia Collaboration, 2) this work, 3) \citet{Carson2013}, 4) \citet{Bonnefoy2014a}, 5) \citet{Hinkley2013}, 6) \citet{Jones2016}.   We conservatively assign a positional uncertainties in each coordinate to account for the difference between the apparent and actual position of the star underneath the coronagraph spot (NIRC2, $\sim$ 0.25 pixels) or from a polynominal fit to the apparent centroid positions derived from satellite spots (CHARIS, $\sim$ 0.25 pixels), uncertainties in the north position angle and pixel scale (larger for CHARIS), the intrinsic signal-to-noise (both), uncertainties in the parallactic angle as recorded in the first header (primarily NIRC2), and uncertainties in the astrometry due to self-subtraction/annealing (both, larger for NIRC2).    The age, gravity, and mass are not directly measured, so we denote their estimates with a "?". }
\label{kapandprop}

\end{deluxetable*}

%% file: charis_astrocal.tex
\begin{deluxetable}{llllll}

\tablecaption{Preliminary SCExAO/CHARIS Astrometric Calibration}
\tiny
\tablehead{{Telescope/Instrument (Coronagraph)} & {UT Date} & {$\rho_{nominal}$ (\arcsec{})} & {PA$_{nominal}$ ($^{o}$)} & {$\rho_{corr}$  (\arcsec{})} & {PA$_{corr}$ ($^{o}$)} }
\startdata
Keck/NIRC2 (Lyot)& 9 December 2017 & 0.784 $\pm$ 0.006 & 244.93 $\pm$ 0.25 & -- & --\\
\hline
SCExAO/CHARIS (Lyot) & 6 September 2017 & 0.797 $\pm$ 0.004 & 242.85 $\pm$ 0.15 & 0.785 $\pm$ 0.008 & 245.05 $\pm$ 0.27\\
SCExAO/CHARIS (SPC) & 6 September 2017 & 0.796 $\pm$ 0.004 & 242.67 $\pm$ 0.13 & 0.784 $\pm$ 0.008 & 244.87 $\pm$ 0.26\\
SCExAO/CHARIS (Lyot) & 4 September 2017 & 0.796 $\pm$ 0.005 & 242.60 $\pm$ 0.30 & 0.784 $\pm$ 0.009 & 244.80 $\pm$ 0.37\\
SCExAO/CHARIS (Lyot) & 16 July 2017 & 0.796 $\pm$ 0.004 & 242.74 $\pm$ 0.15 & 0.784 $\pm$ 0.008 & 244.94 $\pm$ 0.27\\
 \enddata
 \tablecomments{Because our astrometric calibration is focused on the pixel scale and north position angle, we report astrometry in polar coordinates, rather than the usual rectangular coordinates.    The astrometric errors consider variations in centroid measurement (e.g., a simple center-of-light calculation vs. gaussian fitting), the intrinsic signal-to-noise of the detection, and (for the CHARIS corrected astrometry) uncertainties in the absolute pixel scale and true north calibration.  }
\label{charisastromcal}

\end{deluxetable}

%% file: extracted_spectrum.tex
\begin{deluxetable}{lllllll}
\tablecaption{SCExAO/CHARIS Spectrum in Flux Units (Observed)}
\tablehead{\colhead{Wavelength ($\mu m$)} & \colhead{F$_{\nu}$ (mJy)} & \colhead{$\sigma$F$_{\nu}$ (mJy)}} 
\tiny
\centering
\startdata
1.1596 &  0.5910 &  0.0567\\
1.1997 &  0.6112 &  0.0529\\
1.2412 &  0.6942 &  0.0498\\
1.2842 &  0.8349 &  0.0487\\
1.3286 &  0.7658 &  0.0426\\
1.3746 &  0.2586 &  0.0403\\
1.4222 &  0.5371 &  0.0395\\
1.4714 &  0.5830 &  0.0393\\
1.5224 &  0.8103 &  0.0397\\
1.5750 &  0.8248 &  0.0405\\
1.6296 &  1.1274 &  0.0375\\
1.6860 &  1.2744 &  0.0351\\
1.7443 &  1.0190 &  0.0360\\
1.8047 &  0.8616 &  0.0310\\
1.8672 &  0.6344 &  0.0508\\
1.9318 &  0.9750 &  0.0429\\
1.9987 &  0.8215 &  0.0360\\
2.0678 &  1.0233 &   0.0338\\
 2.1394 &  1.2442 &  0.0477\\
2.2135 &  1.3643 &  0.0492\\
2.2901 &  1.2739 &  0.0590\\
2.3693 &  1.2030 &  0.0832\\
 \enddata
 \tablecomments{Spectra are extracted from our conservative ADI/A-LOCI reduction and corrected for throughput losses.   Measurements in regions with non-negligible telluric contamination and/or poor coverage from the template/empirical comparisons ($\lambda$ = 1.3746, 1.4222, 1.4714, 1.8672, 1.9318, and 1.9987 $\mu m$) were not used in our analysis.}
 \label{chariskapandspecapp}
\end{deluxetable}